\documentclass[10pt,a4paper]{article}
\usepackage[english]{babel}
\usepackage{graphicx}

\newcommand{\alt}{\mathbin{\lower 3pt\hbox
   {$\rlap{\raise 5pt\hbox{$\char'074$}}\mathchar"7218$}}}
\newcommand{\agt}{\mathbin{\lower 3pt\hbox
   {$\rlap{\raise 5pt\hbox{$\char'076$}}\mathchar"7218$}}}

\begin{document}
\setcounter{footnote}{0}
\setcounter{equation}{0}
\setcounter{figure}{0}
\setcounter{table}{0}
\vspace*{5mm}

\begin{center}
{\large\bf Conductance of Finite Systems \\ and Scaling in
Localization Theory.
}

\vspace{4mm}
I. M. Suslov \\
P.L.Kapitza Institute for Physical Problems,
\\ 119334 Moscow, Russia \\
E-mail: suslov@kapitza.ras.ru
\end{center}

\begin{center}
\begin{minipage}{135mm}
{\bf Abstract } \\
The conductance of  finite systems plays a central role in the
scaling theory of localization (Abrahams et al, 1979). Usually
it is defined by the Landauer-type formulas, which remain
open the following questions:  (a) exclusion of the contact
resistance in the many-channel case; (b) correspondence of the
Landauer conductance with internal properties of the system; (c)
relation with the diffusion coefficient  $D(\omega,q)$ of
an infinite system.
The answers to these questions are
 obtained below
in the framework of two approaches:  (1) self-consistent theory of
localization by Vollhardt and W$\ddot o$lfle, and (2) quantum
mechanical analysis based on the shell model.  Both approaches
lead to the same definition for the conductance of a finite
system, closely related to the Thouless definition.  In the
framework of the self-consistent theory, the  relations of
finite-size scaling are derived and  the Gell-Mann --
Low functions $\beta(g)$ for space dimensions  $d=1,\,2,\,3$
are calculated.  In
contrast to the previous
attempt by Vollhardt and W$\ddot o$lfle (1982),
the metallic and localized phase are considered
from the same standpoint,
and the conductance of a finite system has no singularity at the
critical point. In the 2D case,
the expansion of  $\beta(g)$ in
$1/g$ coincides  with  results of the $\sigma$ model approach
on the two-loop level and depends on the renormalization
scheme in higher loops; the use of dimensional regularization
 for transition to dimension
$d=2+\epsilon$  looks incompatible with the
physical essence of the problem. The obtained results are
compared with numerical and physical experiments. A situation in
higher dimensions and the conditions for observation of the
localization law $\sigma\propto -i\omega$ for conductivity
are  discussed.  \end{minipage} \end{center}

\twocolumn


\begin{center}
{\bf 1. INTRODUCTION}
\end{center}

\begin{figure}
\centerline{\includegraphics[width=3.5 in]{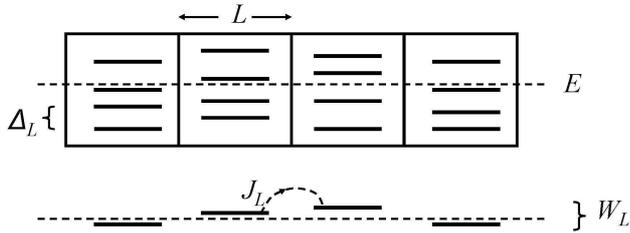}}
\caption{\footnotesize The Thouless scaling construction. The
infinite system is composed of finite blocks of size  $L$;
if only the level closest to the given energy  $E$ is  retained
in each block, the effective Anderson model arises, with
the overlap integral $J_L$ and the scattering of site energies
 $W_L$.} \label{fig1}
\end{figure}

The scaling theory of localization  \cite{1} is based on
consideration of the so called "Thouless number"
$$
g_L=\frac{J_L}{W_L}= \frac{G_L}{e^2/\hbar}  \,,
\eqno(1)
$$
equal to the conductance  $G_L=\sigma_L L^{d-2}$ of the cubic
block of size  $L$ ($\sigma_L$ is the conductivity and  $d$ is a
dimension of space) in units of $e^2/\hbar$, or  to the ratio of
parameters $J_L$ and $W_L$ of the effective Anderson model,
arising in the Thouless scaling construction
(Fig.\,1).  Equivalence of two
representations in (1) follows from the estimate
of overlap integrals  $J_L\sim \hbar/\tau_D$ through the
diffusion time $\tau_D=L^2/D_L$, estimate of $W_L$ as the mean
level spacing $\Delta_L\sim 1/\nu_F L^d$ and the use of the
Einstein relation  $ \sigma_L= e^2 \nu_F D_L $ between
the conductivity $\sigma_L$ and the diffusion constant $D_L$
($\nu_F$ is the density of states at the Fermi level).

The behavior of  $g_L$ at large $L$ is of the main interest:
if  $g_L\to\infty$, then a system is in the metallic phase,
since eigenfunctions of blocks are hybridized with practically
 equal weights; if  $g_L\to 0$, then a system is an Anderson
dielectric (hybridization of the block eigenstates is practically
absent). The block of size $nL$ can be composed  from $n^d$ blocks
of size $L$, so $g_{nL}$ can be recalculated through a given $g_L$
as $g_{nL}=F(g_L,n)$, which for $n\to 1$ can be written in the
differential form
$$
\frac{d \ln g}{d \ln L}= \beta(g) \,,
\eqno(2)
$$
i.e. in the form of the Gell-Mann -- Low equation \cite{2}.
The asymptotic behavior of $\beta(g)$
$$
\beta(g)=\left \{ \begin{array}{cc}
d-2\,,& g \gg 1 \\
\ln g\,, & g \ll 1 \end{array} \right.\,\eqno(3)
$$
follows from the evident relation
$G_L=\sigma_\infty L^{d-2}$  in the metallic phase and the
estimate  $G_L \sim \exp\{-const \cdot L\}$ for a dielectric.
For $d\le 2$, the $\beta$ function is always negative indicating
localization of all states. For $d>2$, it has a root $g_c$,
corresponding to the Anderson transition point
with the power law behavior $\sigma \propto \tau^s$
of the conductivity against the distance $\tau$ to the critical
point.

The qualitative considerations of the paper  \cite{1}
stimulated attempts to formulate them in a more quantitative
form. Conductance of finite systems became a subject of a
vivid discussion \cite{3}--\cite{14} (see a review article
\cite{15})
resulted in establishing of the Landauer approach
\cite{5,12} as an adequate way of description. This approach
reduces the kinetic problem of conductance to the
quantum-mechanical scattering problem.

The original Landauer formula for the strictly 1D
(one-channel) conductor follows from the simple considerations. If
the unit flux of electrons is incident from the left to the
sample under consideration (Fig.\,2,a), then it is transmitted with
\begin{figure}
\centerline{\includegraphics[width=2.5 in]{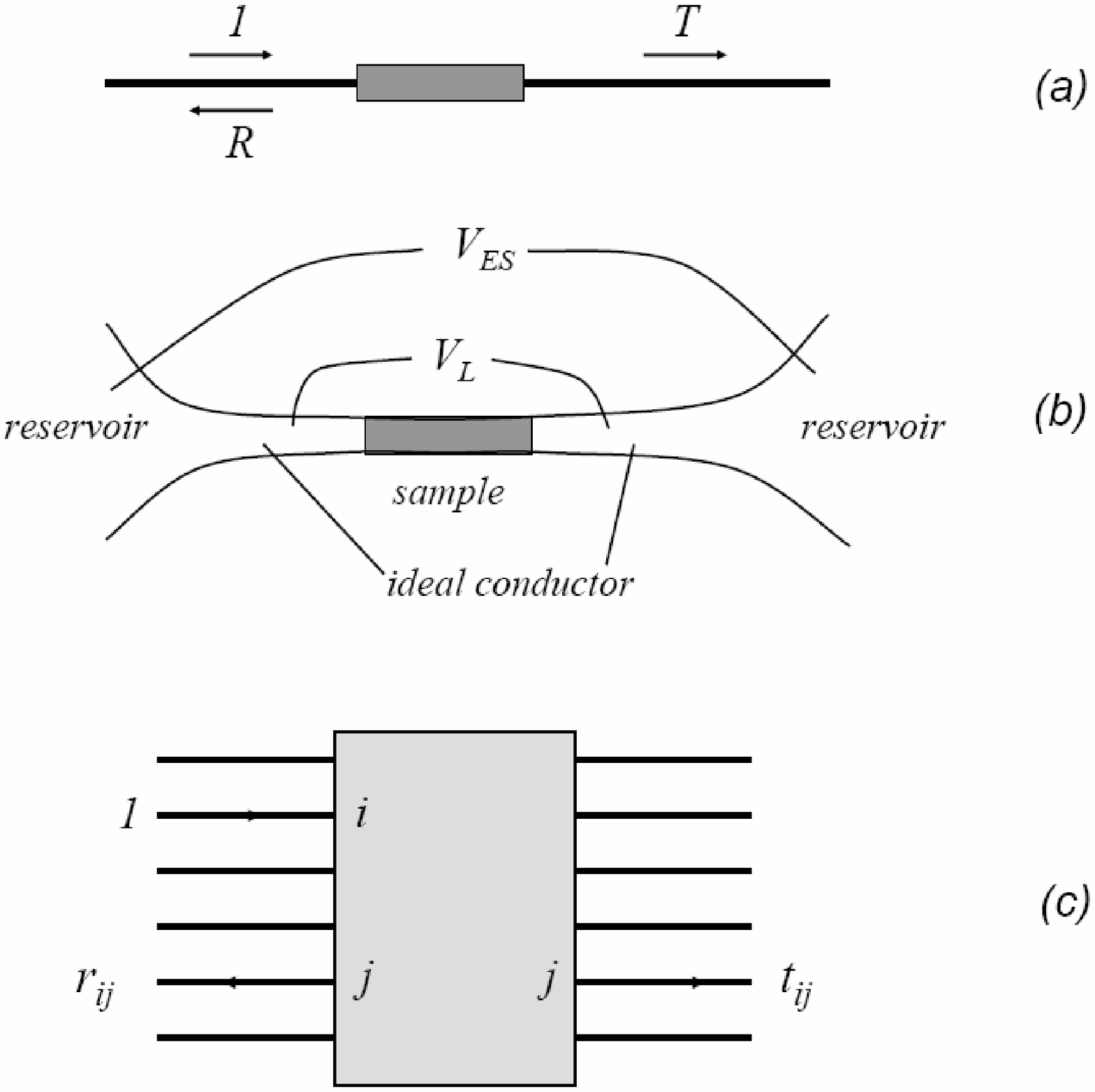}}
\caption{\footnotesize (a) To derivation of the Landauer formula
(4); (b) The difference between Eqs.4 and 5 is determined by the
fact, that voltage  $U_{ES}$ is measured between two reservoirs,
while   voltage $U_L$ between two ideal leads; (c) The
many-channel scattering matrix.} \label{fig2}
\end{figure}
probability  $T$  and
reflected with probability  $R=1-T$
($\,T$ is a transmission coefficient). The current through the
system is proportional to $T$, while the difference of
chemical potentials is determined by the difference of electron
density on the left ($1+R$) and on the right ($T$), i.e.
$1+R-T=2(1-T)$. Consequently the conductance is proportional
to $T/(1-T)$, and estimation of the coefficient gives
\cite{5}\,\footnote{\,Index $L$ at $G_L$ in Eqs.4 and 6
denotes "the Landauer conductance" and not dependence on  $L$.}
$$
G_{L} = \frac{e^2}{2\pi\hbar} \frac{T}{1-T}\,.
\eqno(4)
$$
A somewhat different result was obtained by
Econo- mou and Soukoulis \cite{6} from  linear response
theory
$$
G_{ES} = \frac{e^2}{2\pi\hbar} T\,.
\eqno(5)
$$
Subsequent investigations established
\cite{12,15}, that Eq.4 corresponds to the four-probe, while Eq.5
to the two-probe
measurement geometry  (Fig.\,2,b): the
difference between them is determined by the contact resistance
$2\pi\hbar/e^2$ between the reservoir and the ideal conductor
$$
\frac{1}{G_{L}}= \frac{1}{G_{ES}} - \frac{2\pi\hbar}{e^2}\,.
\eqno(6)
$$
Transition from  one-dimensional  to $d$-dimensional sa- mple
requires consideration of the many-channel scattering matrix shown
at Fig.\,2,c: the plane wave of the unit amplitude incident to the
channel $i$  generates the transmitted and reflected waves with
amplitudes
 $t_{ij}$ and $r_{ij}$  in the channel  $j$. The multi-channel
 generalization of Eq.5 has a form \cite{3,7,11,15}
$$
G_{ES} = \frac{e^2}{2\pi\hbar} \sum\limits_{ij} |t_{ij}|^2\,.
\eqno(7)
$$
Subtraction of the contact resistance in  analogy with  (6)
gives the numerical results  \cite{16} equivalent to the Thouless
definition \cite{17}, relating the conductance with a reaction to
boundary conditions\,\footnote{\,The Thouless definition provides
the  equivalence of two representations (1). The overlap integral
$J_L$ can be estimated as the width of the band occurring from the
given level in the result of the periodic repetition of the block:
it is determined by the change from periodic to antiperiodic
boundary conditions. The Thouless definition is physically
satisfactory but requires consideration of distributions
\cite{17}, being hardly formulated in terms of average quantities.
By this reason it  practically is not used in analytical theory.}.
However, multi-channel generalizations of Eq.4 appear to be
ambiguous \cite{3,8,11,13}, and the problem of  correct exclusion
of the contact resistance remains open.

The previous discussion escapes of the fact that
the conductance of
a finite system is a poorly defined quantity. In a strict
quantum mechanical description, a finite system has a discrete
spectrum and its ground state corresponds to occupation of the
lower levels (Fig.\,3,a)\,\footnote{\,For simplicity, we have in
mind non-interacting electrons in the random potential and
consider one spin projection.}. If the ground state does not
\begin{figure}
\centerline{\includegraphics[width=2.5 in]{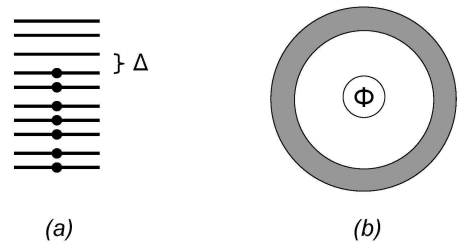}}
\caption{\footnotesize In strictly quantum mechanical description,
a finite system  has no conductance
(a) and no resistance  (b).}
\label{fig3}
\end{figure}
carry a current, then a finite conductance is related with
transitions to excited states, which are separated by a
finite gap $\Delta$; such transitions are absent in the limit of
zero frequency $\omega$, and
$$
{\rm Re} \,G_L(\omega)=0\,,\qquad \omega\to 0 \,.
\eqno(8)
$$
It is curious, that in the Aharonov--Bohm geometry (Fig.\,3,b) the
ground state carries a current, if the magnetic flux $\phi$
through the ring-shaped sample is not equal to the integer or
semi-integer number of quanta $\phi_0=\hbar c/e$ \cite{14}; in
this case, a persistent current flows through the system without
external voltage and the resistance
 $R_L$ is also zero,
$$
{\rm Re}\,R_L(\omega)\to 0\,,\qquad \omega\to 0 \,
\eqno(9)
$$
(contradiction with Eq.8 is avoided due to the presence of
${\rm Im}\,G_L(\omega)$).

In fact, this problem is well-known: the formulas of  linear
response theory are complemented by a prescription that the
entering them $\delta$ functions should be smeared out by the
quantity $\gamma$,
next the thermodynamic limit $L\to\infty$ is taken and
only then  $\gamma$ is set to zero.
In fact, such procedure transforms the discrete spectrum into
the continuous density of states (Fig.\,4). For a finite system
such a procedure becomes impossible and attenuation $\gamma$
should remain finite. The question arises on the origin of
this attenuation and its dependence on parameters.
\begin{figure}
\centerline{\includegraphics[width=3.1 in]{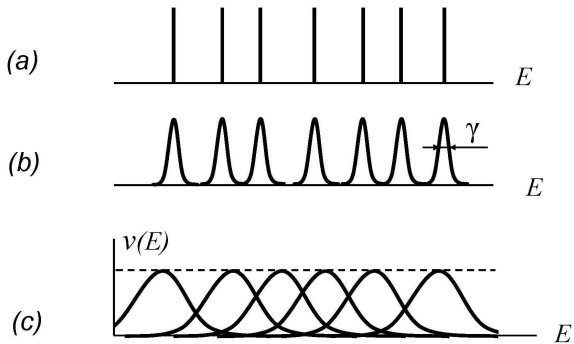}}
\caption{\footnotesize The discrete levels (a) acquire
width  $\gamma$ due to a finite lifetime (b).
If the limit  $L\to\infty$ is taken before the limit
$\gamma\to 0$, the extended levels
overlap strongly
and form the continuous density of states (c).} \label{fig4}
\end{figure}

Attempts of discussing
this question were made in
the papers \cite{18}--\cite{21}, based on the "shell model"
developed in nuclear physics for description of coupling
between  a
discrete spectrum of the "target" and a continuous spectrum of
scattered particles \cite{210}. Unfortunately, the physics
of the problem
remain unclear in these papers, because the
$\sigma$ model formalism was introduced at the early
stage of consideration.
In addition,
the bare diffusion
coefficient was considered as a given constant, while it
actually  depends on the degree of the openness of a system.

In derivation of the Landauer formulas \cite{6,7,8,15} the
indicated problem is avoided in a following manner. The system
under consideration is connected with the ideal leads, which can
be taken  sufficiently massive;  the spectrum  becomes
quasi-continu- ous and attenuation  $\gamma$ can be tended to
zero. Therefore, the Landauer conductance corresponds to the
composite system "sample+external leads" and
 not to the system under consideration.
 This point is especially clear from the
fact that the matrix elements entering the Kubo formula are
determined by integration over the region of ideal leads.

The question arises, in what extent the formulas (4--7)
reflect the internal properties of the
system.
To illustrate it more clearly, let  introduce
the potential barrier between the sample and ideal leads.
If the height of the barrier tends to infinity, then
the Landauer  resistance grows unboundedly, while
nothing occurs with the system
itself. Contrary,
if the height of the barrier  tends to zero,
then the boundary resistance disappears but the system is
dangerously affected by its environment.

There is one more question. The infinite system is fully
characterized by the diffusion coefficient  $D(\omega, q)$,
which generally
possesses
the temporal and spatial dispersion.
The conductance of a finite system is evidently  related to
$D(\omega, q)$ but this relation is not clear in the Landauer
approach.

Therefore, the following points remain unclear at the
present time:

(a) exclusion of the reservoir contact resistance in the
many-channel case;

(b) relation of Eqs.4,5,7 with internal properties of the
system;

(c) relation of the Landauer conductance with the diffusion
coefficient  $D(\omega, q)$ of an infinite system.

The answers to these questions are obtained below in the framework
of two approaches: (1) self-consistent theory of localization by
Vollhardt and W$\ddot o$lfle \cite{22,23}, and (2)
quantum-mechanical analysis based on the shell model
\cite{18,210}. Both approaches lead to the same definition for the
conductance of a finite system, closely related to the Thouless
definition: it gives a new  strong argument in favour of the
self-consistent theory. Further, in the framework of this theory,
we  calculate the Gell-Mann -- Low functions  $\beta(g)$ for the
space dimensions $d=1,\,2,\,3$ (Fig.\,5). In contrast to the
analogous calculation by Vollhardt and W$\ddot o$lfle \cite{23},
the $\beta$ function has no singularity in the fixed point $g_c$.
The latter is related with the fact that the metallic and
localized phase are considered from the same standpoint, so  the
conductance of a finite system has no singularity at the critical
point: it is in agreement with the general principles of the
modern theory of critical phenomena \cite{24,25}.

The present theory has the following structure.
A finite system is topologically quasi-zero-dimensional and its
effective dimensionality is less than two. All states of
this system are formally localized and one can introduce
the finite correlation length $\xi_{0D}$; it  satisfies
the scaling relation
$$
\frac{\xi_{0D}}{L}
=F\left( \frac{L}{\xi}\right)\,,
\eqno(10)
$$
analogous to that for quasi-one-dimensional systems
\cite{266,26,27,28}; $\xi$ is the correlation length of the
infinite $d$-dimensional system. The diffusion coefficient
has a behavior typical for the dielectric phase,
$$
D(\omega, 0) = (-i\omega) \xi_{0D}^2 \,,
\eqno(11)
$$
and turns to zero in accordance with  (8,9). The above statements
are valid  only for closed systems.  In  open systems the finite
diffusion coefficient $D_L$ arises, and the following result can
be derived for the
dimensionless conductance
$$
g_L =F_1\left( \frac{\xi_{0D}}{L}\right)  \,.
\eqno(12)
$$
Replacement of $\ln L$ by $\ln(L/\xi)$ in the Gell-Mann
-- Low equation  (2) allows to represent  $g_L$ as a
function of  $L/\xi$. The latter  can be determined
from (10,\,12) and allows to reconstruct  $\beta(g)$.
\begin{figure}
\centerline{\includegraphics[width=3.1 in]{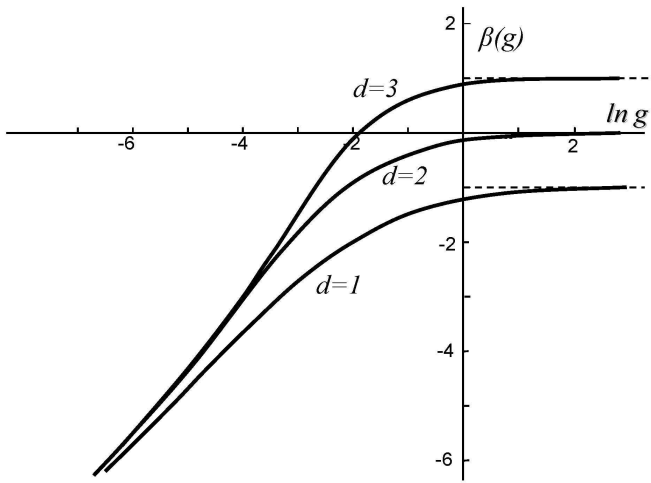}}
\caption{\footnotesize Gell-Mann -- Low functions  $\beta(g)$
for $d=1,\,2,\,3$ obtained in the present paper.}
\label{fig5}
\end{figure}

The paper is organized  as follows. In Sec.\,2 we discuss the
early attempt of  scaling by Vollhardt and W$\ddot o$lfle
\cite{23} and formulate the main difference between it and the
present paper. In Sec.\,3 the correlation length  $\xi_{0D}$ is
discussed and the scaling relation (10) is derived for $d<4$. In
Sec.\,4 we consider open systems and derive the result of type
(12). The same result is derived in Sec.\,5 from the shell model;
its physical sense is clarified and dependence  on the measurement
geometry is discussed. In Sec.\,6 the length dependence of $g_L$
is presented in the clear form and $\beta$ functions for
$d=1,\,2,\,3$ are calculated. Their expansion in powers of $1/g$
is compared with the results of the $\sigma$ model approach
\cite{31,32}; the use of dimensional regularization in $\sigma$
models is found to be in conflict with the physical essence of the
problem. In Sec.\,7 the obtained results are compared with
numerical
 \cite{29,30} and physical \cite{211,49} experiments.  A
 situation in higher dimensions $d\ge 4$ is discussed in Sec.\,8.
In Sec.\,9 we summarize and discuss
 the consequences of the present study for the
 conductance distribution, spatial dispersion of the
 diffusion coefficient and  observation
 of the localization behavior $\sigma(\omega)\propto -i\omega$.

\vspace{2mm}

\begin{center}
{\bf 2. SELF-CONSISTENT THEORY AND SCALING} \end{center}

Self-consistent theory of localization by  Vollhardt and
W${\rm {\ddot o}}$lfle theory is based on existence of the
diffusion pole in the irreducible four-leg vertex  $U_{{\bf k}
{\bf k}^\prime} ({\bf q})$,
entering the Bethe--Salpeter equation and playing the
role of the scattering probability  $W_{{\bf k}
{\bf k}^\prime}$ in the quantum kinetic equation.
Using the estimate in the spirit
of $\tau$-approximation, $D \propto
\langle U \rangle ^{-1}$, where   $\langle ... \rangle$
is  averaging over momenta,
one can obtain the
self-consistency equation  \cite{22,23}
which can be presented in the form
\cite{28}\,\footnote{\,Equation of type (13) can be obtained by
approximate solution of the Bethe-Salpeter equation \cite{22}
or by detailed analysis of spectral properties of the quantum
collision operator  \cite{33}. The possibility to neglect the
spatial dispersion  of the diffusion coefficient was justified in
 \cite{33}. }
$$
\frac{E^2}{W^2} = \frac{D(\omega)}{D_{min}} + \Lambda^{2-d}
\int\limits_0^\Lambda \frac{d^dq}{(2\pi)^d}
\,\frac{1}{[-i\omega/D(\omega)] + q^2} \,
\eqno(13)
$$
(the limits of integration are written for the modulus of ${ q}$).
Here $E$ is the energy of the bandwidth order,  $W$ is the
amplitude of disorder, $\Lambda$ is the ultraviolet cut-off,
$D_{min}$  is a characteristic scale of the diffusion
constant corresponding to the Mott minimal
conductivity.

The metallic phase is possible for  $d>2$, when a value of
 the basic integral
$$
I(m) = \int\limits_0^\Lambda \frac{d^dq}{(2\pi)^d}
\,\frac{1}{m^2 + q^2}   \,
\eqno(14)
$$
is finite for $m=0$.
Accepting $D=const>0$ for $\omega\to 0$
and specifying $\tau$ as a distance to a transition, one has
$$
D=D_{min}\, \tau\,,\qquad
\tau = \frac{E^2}{W^2} -I(0) \Lambda^{2-d} \,,
 \eqno(15)
 $$
so the exponent of conductivity is unity.
In the dielectric phase one makes substitution
$D= -i\omega \xi^2$, and Eq.13 determines
 the correlation length $\xi$; in particular,
 for  $d>2$
$$
\xi\sim a\, |\tau|^{-\nu}\,,
\qquad
$$
$$
{\rm где} \qquad
\nu=\left \{ \begin{array}{cc}
1/(d-2)\,,& 2<d<4 \\
1/2\,, & d>4 \end{array} \right.\,.
\eqno(16)
$$

The attempt of using Eq.13 for derivation of scaling
equations was made in \cite{23} and
contains two ingredients.
\vspace{2mm}

{\it 1. Modification of the Einstein relation.  }
According to  \cite{23}, the Einstein relation is
modified in the localized phase due to non-local
effects and acquires the additional exponential factor
$$
\sigma_L\,\sim\, e^2 \nu_F D_L \, e^{-L/\xi} \,.
\eqno(17)
$$
To obtain this result, one considers the change of the electron
density  $\rho(x)$, induced by the scalar potential
 $\varphi(x)$,
$$
\rho(x)= \int\limits_{-L/2}^{L/2} \alpha(x-x')
                       \varphi(x') dx'\,,
\eqno(18)
$$
where  $\alpha(x-x')$  is polarizability
$$
\alpha(x-x')= - e^2 \nu_F \left[ \delta(x-x') - \right.
$$
$$
        \left.  -(2\xi)^{-1} \exp\{-|x-x'|/\xi\} \right] \,.
\eqno(19)
$$
For a closed system,
the diffusion current $j_{diff}(x)=-D_L d\rho(x)/dx$
at the boundaries of the system $x=\pm L/2$
is compensated by the electric current
$j_e(x)=\sigma_L E$, which  allows
to determine  $\sigma_L$. Producing such calculations for
$$
\varphi(x) = \varphi_0 -Ex \,,
\eqno(20)
$$
one has
$$
\rho(x)= e^2 \nu_F \left[\,
E(L/2+\xi) e^{-L/2\xi}\sinh(x/\xi)- \right.
$$
$$ \left.
        -\varphi_0 e^{-L/2\xi}\cosh(x/\xi)
\,\right]  \,,
\eqno(21)
$$
and
$$
j_e(\pm L/2)= e^2 \nu_F D_L
    \left[\,\frac{ E(L/2+\xi)\mp \varphi_0 }{2\xi}\,+
\right.
$$
$$
 \left.       + \,\frac{  E(L/2+\xi)\pm\varphi_0}{2\xi}
               e^{-L/\xi}\,\right]
\eqno(22)
$$
Accepting  $\varphi_0=\pm E\left(L/2+\xi\right)$, one
obtains
$$
j_e(\pm L/2)= e^2 \nu_F D_L
         \left(\,1+L/2\xi\,\right) e^{-L/\xi} \cdot E
\eqno(23)
$$
in accordance with (16). It is easy to see that this result is
related with the unphysical response  to the constant potential
$\varphi_0$, which is  a consequence of accepted
approximations; absence of self-consistency is especially clear
for $\varphi_0\ne 0$, when estimation of
$\sigma_L$ using  $j_e(L/2)$ and $j_e(-L/2)$ gives  different
results.\,\footnote{\,In fact, the kernel $\alpha(x,x')$ should be
constructed as  a binary expansion in eigenfunctions of the
diffusion operator; for open systems, integration in  (18) should
be taken over the whole space.} Of course, the correct
consideration recovers validity of the Einstein relation. In the
framework of the self-consistent theory, $\sigma(\omega,q)$ and
$D(\omega,q)$ are effectively independent of $q$ \cite{33},
providing the local response in the coordinate space;
the relation between them has a local character and cannot be
modified due to restriction of the system size. The absence of
the factor $\exp\{-L/\xi\}$ is catastrophic for the paper
\cite{23}, since it fails to obtain  the result $g_L\sim
\exp\{-const\cdot L\}$ in the localized phase.

\vspace{2mm}
{\it 2. Modification of the self-consistency equation.  }
For a finite system, equation (13) is modified by introducing
the lower  cut-off,
$$
\frac{E^2}{W^2} =
\frac{D_L}{D_{min}} + \Lambda^{2-d} \int\limits_{\sim
1/L}^\Lambda \frac{d^dq}{(2\pi)^d} \,\frac{1}{m^2 + q^2} \,,
\eqno(24)
$$
and rearranged by subtraction of the same equation with
$L=\infty$:
$$
D_L=D_\infty + D_{min}\Lambda^{2-d}
\int\limits_0^{\sim 1/L} \frac{d^dq}{(2\pi)^d}
\,\frac{1}{m^2 + q^2} \,,
$$
$$
\qquad m=\xi^{-1}    \,.
\eqno(25)
$$
Since  $D_\infty\sim \tau$ in the metal and
$D_\infty=0$ in the dielectric phase, the diffusion
coefficient of a finite system
acquires a singularity
at the critical point. It is in conflict with  general
principles of the  modern theory of critical phenomena
\cite{24,25},  which allow a phase transition  only
in the thermodynamic limit  $L\to\infty$.

The present theory is also based on Eq.13, while indicated defects
are removed in the following manner. A finite system is
topologically zero-dimensional and all its states are formally
localized, though the effective correlation length $\xi_{0D}$
coincides with $\xi$ only in the deep of the localized phase
(in the metallic regime,  $\xi_{0D} > L$).
As a result, $D_\infty$ turns
to zero  in both phases and Eq.25 becomes almost
satisfactory. The correct result is obtained below,
$$
D_L= D_{min}\Lambda^{2-d} \cdot \frac{1}{L^d}
\sum\limits_{\rm q} \left.\,\frac{e^{i\rm q\cdot x}}{m^2 + q^2}
\right|_{|{\rm x}|\sim L} \,,
$$
$$\qquad m=(\xi_{0D})^{-1}
\eqno(26)
$$
and differs from (25) by replacement of the integral by
the discrete sum and
concretization of the way of cut-off:
the latter is provided by the oscillating factor $e^{i \rm
q\cdot x}$, which effectively restricts
summation by values  $|{\bf q}|\alt 1/L$.  These
modifications are crucial for derivation of the result
$D_L\sim \exp\{-L/\xi\}$ in the localized phase.

The application of the quasi-zero-dimensional concept is not a
pure theoretical construction, but allows to distinguish the
real behavior of open and closed systems. Absence of
singularities in small systems can be also verified
experimentally. Therefore, the present theory differs from
\cite{23} on the level of observable consequences.

\vspace{4mm}
\begin{center}
{\bf 3. CORRELATION LENGTH OF
QUASI-ZERO-DIMENSIONAL SYSTEM} \end{center}

\begin{center}
{\bf 3.1. Dimensions $2<d<4$}
\end{center}

A finite system is considered as quasi-zero-dimen- sional and
its correlation length  $\xi_{0D}$ can be studied in
analogy with the quasi-one-dimensional case  \cite{28}.
In a finite system, the basic integral (14) is replaced by
the discrete sum
$$
I(m) = \frac{1}{L^{d}} \sum\limits_{|{\rm q}|<\Lambda}
\,\frac{1}{m^2 + q^2}  \,,\qquad
m^{-1}=\xi_{0D} \,,
\eqno(27)
$$
where allowed values of  ${\bf q}$ has a form $2\pi{\bf s}/L$,
and ${\bf s}=(s_1,s_2,\ldots,s_d)$ is the $d$-dimensional
vector with integer components $s_i=0,\,\pm 1,\,\pm 2,\,\ldots$
We accept the periodic boundary conditions in all directions,
which correspond to the closed system  (Sec.\,4).
The term with  ${\bf q}=0$ provides the divergency of $I(m)$ at
$m\to 0$ and the system is always in the localized
regime. It is convenient to make the following
decomposition
$$
I(m) = \frac{1}{ L^{d}}
\,\frac{1}{ m^2} + \frac{1}{ L^{d}} \sum\limits_{
\begin{array}{c} { \scriptstyle {\rm q}\ne 0 } \\
{\scriptstyle |{\rm q}|<\Lambda  } \end{array}
 }
\left( \,\frac{1}{m^2 + q^2} -\,\frac{1}{ q^2}\right)\,
$$
$$
+\,\frac{1}{L^{d}} \sum\limits_{
\begin{array}{c} { \scriptstyle {\rm q}\ne 0 } \\
{\scriptstyle |{\rm q}|<\Lambda  } \end{array}
}
 \frac{1}{q^2 } \, \,\equiv \,
I_1(m)+I_2(m)+I_3(0)
\,,
\eqno(28)
$$
where we separated the term with ${\bf q}=0$, and the rest
of the sum
is rearranged by addition and subtraction of the analogous sum
with $m=0$. The limit $\Lambda\to\infty$ can be taken  in the
second term $I_2(m)$ transforming it to the form $L^{2-d} H_0(mL)$
(neglecting contributions $\sim m^2 \Lambda^{d-4}$).
The third term $I_3(0)$ can be calculated at  $L\to\infty$
by the change of summation by integration, while for
finite $L$ it has a structure\,\footnote{\,It can be obtained
using the $\alpha$-representation (see  Appendix) with the
cut-off $|q_i|<\Lambda$.}
$$
I_3(0) = \Lambda^{d-2}  \left\{ b_0
+ b_1 \left(\frac{a}{L} \right)^{d-2}+
\right.
$$
$$ \left.
+ b_2 \left(\frac{a}{L} \right)
+ b_3 \left(\frac{a}{L} \right)^2
+\ldots \right\}   \,,
\eqno(29)
$$
where we accepted $a=\Lambda^{-1}$. Substitution of
(28,\,29) into the self-consistency equation  (13) gives
$$
\left(\frac{L}{a} \right)^{d-2}
\left[\tau + O(m^2 a^2)
+ O \left(\frac{a}{L} \right) \right]
=
$$
$$ =b_1 + H_0(mL) +\frac{1}{(mL)^2}  \,,
\eqno(30)
$$
where definition  $ \tau = E^2/W^2 - b_0$  coincides with
(15), since $b_0$ corresponds to the value $I(0)$,
calculated in the integral approximation. According to (30),
$\xi_{0D}$ is a regular function of $\tau$.
Expressing $\tau$ through the correlation length $\xi$ of the
$d$-dimensional system  ($\xi^{-1/\nu} \sim |\tau|=\pm \tau$) and
omitting terms vanishing at $a\to 0$, one has
$$
\pm c_d \left(\frac{L}{\xi} \right)^{d-2}
= H \left(\frac{L}{\xi_{0D}} \right) \,,
\eqno(31)
$$
$$
H(z)=\frac{1}{4 \pi^2} \sum\limits_{ {\bf s} \ne 0 }
\left( \,\frac{1}{{|{\bf s}|^2 + (z/2\pi)^2}}
-\,\frac{1}{ |{\bf s}|^2} \,
\right)\,+
$$
$$
+b_1+\,\frac{1}{z^2}    \,,
\eqno(32)
$$
which is the desired scaling relation (10),
consisting of two branches
($c_d$ are positive coefficients introduced in  \cite{28}).
The asymptotical behavior of  $H(z)$
$$
H(z)= \left \{ \begin{array}{cc}
\displaystyle{1/z^2}\,,& \qquad z\ll 1 \\
-A(z-z^*)\,,& \qquad z\to z^* \\
-c_d z^{d-2}\,,& \qquad z\gg 1
\end{array} \right.  \,
\eqno(33)
$$
is obtained noticing that $H(z)$ at small  $z$
is determined by the last term in (32), while for large $z$
the sum over  ${\bf s}$ is approximated by the integral; the
regular expansion is possible near the root  $z^*$, corresponding
to the critical point. At arbitrary $z$ the sum over $\bf s$
can be calculated numerically, giving   $H(z)$
for $d=3$ shown in Fig.\,6. Introducing the variables
$$
y=\xi_{0D}/L\,,\qquad x=\xi/L  \,,
\eqno(34)
$$
one has for dependence  $y(x)$  (Fig.\,7)
\begin{figure}
\centerline{\includegraphics[width=3.1 in]{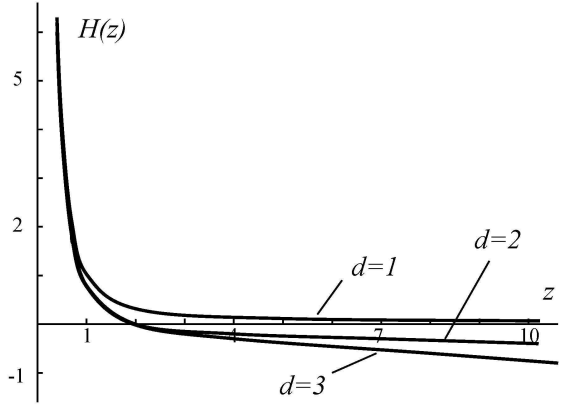}}
\caption{\footnotesize Functions  $H(z)$ for $d=1,\,2,\,3$.}
\label{fig6}
\end{figure}
\begin{figure}
\centerline{\includegraphics[width=3.1 in]{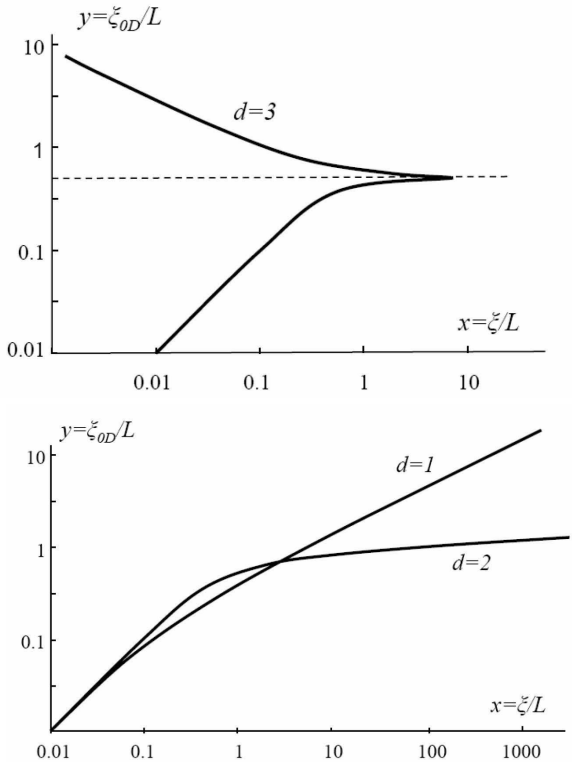}}
\caption{\footnotesize Behavior of $\xi_{0D}/L$ versus
$\xi/L$ for $d=1,\,2,\,3$.} \label{fig7}
\end{figure}
$$
y=
\left \{ \begin{array}{cc}
\left( c_d/x^{d-2}\right)^{1/2} \,,& \qquad y\gg 1 \\
y^* \pm const/ x^{d-2} \,,& \qquad y\to y^*\\
x \,,& \qquad y\ll 1
\end{array} \right.  \,.
\eqno(35)
$$
The constant $b_1$ in (29) is not universal and depends on the way
of cut-off: a value $b_1=- 0.226$   for a spherical cut-off
($|q|<\Lambda$) is used below for $d=3$, though in general it
should be considered as an adjustable parameter.\,\footnote{\,For
the cubical cut-off ($|q_i|<\Lambda$) one has $b_1=- 0.0314$.  }
Corrections to
scaling can be obtained from  (30) and have the same form as for
quasi-one-dimensional systems: it confirms universality of
their structure argued  in \cite{28}.

\begin{center}
{\bf 3.2. Two-dimensional case}
\end{center}

In two dimensions we have
$$
I_3(0) = \frac{1}{2\pi} \ln\frac{L}{a}\, +
 b_1 +\ldots ,\quad
\frac{E^2}{W^2} = \frac{1}{2\pi} \ln\frac{\xi}{a} ,
\eqno(36)
$$
and the scaling relation has the form
$$
\frac{1}{2\pi}\ln\left(\frac{\xi}{L} \right)
= H \left(\frac{L}{\xi_{0D}} \right)
\eqno(37)
$$
with the previous definition of $H(z)$
(Fig.\,6). Using the asymptotic results
$$
H(z)=
\left \{ \begin{array}{cc}
\displaystyle{1/z^2}\,,& \qquad z\ll 1 \\
-(1/2\pi)\ln z\,,& \qquad z\gg 1
\end{array} \right.  \,, \qquad
\eqno(38)
$$
we have  $y=x$ for  $x\ll 1$ and $y\sim \left[\ln x
\right]^{1/2}$ for  $x\gg 1$  (Fig.\,7).
Below we use the value  $b_1=0.1780$ obtained for
the spherical cut-off.

\begin{center}
{\bf 3.3. Dimensions $d<2$}
\end{center}

For $d<2$, subtraction of the  term with  $m=0$ is
not necessary and the limit  $\Lambda\to\infty$ can be taken
immediately in  (27). The scaling relation has the form
$$
c_d \left(\frac{L}{\xi} \right)^{d-2}
= H \left(\frac{L}{\xi_{0D}} \right) \,,
$$
$$ H(z)=
 \frac{1}{(2\pi)^2} \sum\limits_{ \bf s }
\left( \,\frac{1}{ (z/2\pi)^2 +|{\bf s}|^2 }
\,
\right)\,
\eqno(39)
$$
and consists of one branch, since the function  $H(z)$
is positive (Fig.\,6). Its asymptotic behavior
$$
H(z)=
\left \{ \begin{array}{cc}
\displaystyle{1/z^2}\,,& \qquad z\ll 1 \\
c_d/ z^{2-d}\,,& \qquad z\gg 1
\end{array} \right.  \,,
\eqno(40)
$$
is obtained analogously and and gives
$y=x$ for $x\ll 1$ and $y\sim x^{(2-d)/2}$
for $x\gg 1$  (Fig.\,7).
In the 1D case the function  $H(z)$ can be calculated
exactly (see Eq.118 below).

\vspace{4mm}

\begin{center}
{\bf 4. SITUATION IN OPEN SYSTEMS}
\end{center}

\begin{center}
{\bf 4.1. Difference between open and closed systems}
\end{center}

Consider the effective diffusion equation
$$
\frac{\partial f}{\partial t}-
D\, \nabla^2 f = 0\,,
\eqno(41)
$$
describing the electron distribution $f({\bf x},t)$.  In an
infinite system the operator $-\nabla^2$ has eigenfunctions
$e_s({\bf x})=e^{i \rm q_s \cdot x}$ and eigenvalues
$\lambda_s=q_s^2$. In a finite system
$e_s({\bf x})$ and $\lambda_s$  become non-trivial and
determine the evolution  of the initial distribution
$f_0({\bf x})$
$$
f({\bf x},t)=
\sum_s A_s e^{-D\lambda_s t} e_s({\bf x})\,, \qquad A_s=(f_0,e_s)\,.
\eqno(42)
$$
The difference between open and closed systems can be
formulated on a very abstract level: in the first case,
the minimal eigenvalue  $\lambda_0$ is zero and corresponds
to the constant eigenfunction
$$
\lambda_0=0 \,,\qquad e_0({\bf x})=const\,,
\eqno(43)
$$
while in the second case
$$
\lambda_0>0\,,\qquad e_0({\bf x})\ne const\,.
\eqno(44)
$$
In the first case one has from (42) at  $t\to\infty$
$$
\left. f({\bf x},t)\right|_{t\to\infty}=
const =\langle f_0 \rangle
\,, \eqno(45)
$$
i.e. distribution  $f({\bf x},t)$ tends to the constant limit,
equal to the average value of  $f_0({\bf x})$
in the coordinate space;
consequently, environment does not affect the natural
process of relaxation, and the total number of particles is
conserved. In the second case
$$
\left. f({\bf x},t)\right|_{t\to\infty}=
A_0 \, e_0({\bf x})\, e^{-D\lambda_0 t}\,,
 \eqno(46)
$$
i.e. distribution over ${\bf x}$ is stabilized, but its
amplitude is decreased due to escape of the particles
through boundaries of the system; by the same reason their
density near boundaries is less than in the center.

Consider examples. Let the  1D system is arranged in the
interval $0\le x \le L$.  For  boundary conditions of
the Bloch  type
$$
 f(L)= f(0) e^{ i\varphi}\,
 \eqno(47)
$$
the allowed values of $q$ has a form $q_s=(2\pi s+\varphi)/L$
with integer  $s$, so the system is closed at $\varphi=0$, and
maximally open at $\varphi=\pi$, i.e. for periodic and
antiperiodic conditions correspondingly. For the more realistic
boundary conditions
$$
f'_x(0)=\kappa f(0)\,, \qquad
f'_x(L)=-\kappa f(L)\,
 \eqno(48)
$$
the system is closed for   $\kappa=0$ due to the absence of flow
through boundaries; the maximum openness is realized at
$\kappa=\infty$, i.e. for the zero boundary conditions.

\begin{center}
{\bf 4.2. Attenuation of electron states and finiteness
of the diffusion coefficient} \end{center}

In the open system,  electrons can escape through the
boundaries and their eigenstates has a finite lifetime.
It can be proved quite generally, that it provides
the finite diffusion constant in the static
limit.

Consider the density correlator
$$
{\cal K}_\omega({\bf r-r'})=\langle
           G_{E+\omega}^R({\bf r,r'})
                G_{E}^A({\bf r',r})\rangle   \,,
\eqno(49)
$$
where $G^R$ and $G^A$ are the retarded and advanced Green
functions. Eq.49 can be rewritten identically
$$
{\cal K}_\omega({\bf r-r'})=\int_{-\infty}^{\infty} d\epsilon
              \int_{-\infty}^{\infty} d\omega' \,
   \frac{1}{E+\omega-\epsilon+i0}\,\cdot
$$
$$
\qquad\cdot\, \frac{1}{E-\omega'-\epsilon-i0} \,
         \rho_{\epsilon,\epsilon+\omega'}({\bf r-r'})\,,
\eqno(50)
$$
where  $\rho_{\epsilon,\epsilon+\omega}({\bf r-r'})$ is
the Berezinsky -- Gor{$'$}kov spectral density, whose Fourier
transform is related with polarizability $\alpha(\omega,q)$
\cite{33}
$$
\rho_{\epsilon,\epsilon+\omega}(q)=-
   \frac{{\rm Im}\,\alpha_\epsilon(\omega,q)}{\pi e^2 \omega}\,.
\eqno(51)
$$
Using the definitions of kinetic coefficients and analytic
properties of the response functions  \cite{34}, it is easy to
show that \cite{33}
$$
{\cal K}_\omega(q)=   \frac{2\pi\nu_{F}}
{-i\omega+D(\omega,q) q^2}  \,,
\eqno(52)
$$
where $D(\omega,q)$ is the observable diffusion coefficient. It
should be stressed, that this result has a general character and
is not restricted by the metallic phase.

The finite lifetime of the electron states leads to a replacement
of infinitesimal quantities $\pm i 0$, entering definitions of the
retarded and advanced Green functions, by $\pm i\gamma$; analogous
changes occur in (50). Reproducing the indicated calculations, we
come to conclusion that the replacement
$$
-i\omega \,\,\longrightarrow\,\, -i\omega + 2\gamma \,
\eqno(53)
$$
should be made in (52), both in the term $-i\omega$ and
in $D(\omega,q)$. In the localized phase the following
combination remains invariant
$$
\frac{-i\omega}{D(\omega,q)} =
\frac{-i\omega}{(-i\omega) \xi^2}\, \,\,\longrightarrow \,\,\,
\frac{-i\omega+2\gamma}{(-i\omega+2\gamma) \xi^2} \,,
\eqno(54)
$$
which has a simple physical sense: attenuation of eigenstates was
introduced for the permanent eigenfunctions, so the correlation
length $\xi$ characterizing the latter is also unchanged. In the
static limit $D(\omega,q)$ is replaced by $ D_L=2\gamma \xi^2$,
i.e. the finite diffusion constant arises.

\begin{center}
{\bf 4.3. Modification of the self-consistency equation}
\end{center}

The result  $D(\omega)\to 0$  (Sec.\,3) is valid for
closed systems, being directly related with the existence
of the allowed value  ${\bf q}=0$; the self-consistency equation
(13) has a form
 $$
\frac{E^2}{W^2} =  \Lambda^{2-d}
\cdot\frac{1}{L^{d}} \sum^{(c)}_{\rm q}
\,\frac{1}{m^2 + q^2}  \,,\qquad
m^{-1}=\xi_{0D} \,.
\eqno(55)
$$
In the open system, the diffusion constant  $D_L$ becomes
finite, but  the
correlation length remains unchanged:
$$
\frac{E^2}{W^2} = \frac{D_L}{D_{min}} + \Lambda^{2-d}
\cdot\frac{1}{L^{d}} \sum^{(o)}_{\rm q}
\,\frac{1}{m^2 + q^2}  \,.
\eqno(56)
$$
The labels $(c)$ and  $(o)$ indicate the closed and open
system, which have the different sets of allowed values of
${\bf q}$ in the sum. Taking the difference of (55) and (56),
one obtains
$$
D_L=D_{min} \Lambda^{2-d}
\cdot\frac{1}{L^{d}} \left(
\sum^{(c)}_{\rm q} \,\frac{1}{m^2 + q^2}
-\sum^{(o)}_{\rm q} \,\frac{1}{m^2 + q^2}
                        \right) \,,
\eqno(57)
$$
which can be considered as a definition of the diffusion
coefficient for a finite system. This definition contains
essential freedom, since the choice of the open and closed system
is a subject of agreement.
For the Bloch boundary
conditions\,\footnote{\,In general, the factor $L^{-d}$ before
the sum over  ${\bf q}$ can be replaced by the more complicated
normalization factor (see Eq.108 below), which can give
the power corrections in  $1/L$ (if treated inaccurately),
and destroy the exponent in (59). Such problems are
absent for eigenfunctions in the form of plane waves,
corresponding to the Bloch conditions (47).  The
realistic boundary conditions (48) are considered in
Sec.5.7.} (47)
one can accept as
etalons the systems with $\varphi=0$ and $\varphi=\pi$,
$$
g_L= L^{d-2} \cdot\frac{1}{L^{d}}
\left(
\sum^{(\varphi=0)}_{\rm q} \,\frac{1}{m^2 + q^2}
-\sum^{(\varphi=\pi)}_{\rm q} \,\frac{1}{m^2 + q^2}
                        \right) \,,
\eqno(58)
$$
where we came from  $D_L$ to $g_L$, accepting
$D_{min} \Lambda^{2-d} =1/\hbar \nu_F$. We shall refer
Eq.58 as the "Thouless definition", since  $g_L$ is
determined by the change from periodic to antiperiodic
boundary conditions\,\footnote{\,Strictly
speaking, the original Thouless definition deals with
the boundary conditions for the electron wave function,
and not for the effective diffusion problem.
Probably, it is the most close correspondence that can be
established in terms of the averaged quantities (see Footnote
2).  }.  We accept  such a change along only one of coordinate
axes, remaining periodic conditions in other directions:
according to Sec.\,5, it corresponds to the natural experimental
geometry.

The definition  (58) provides the exponential decrease of
$g_L$ in the localized phase (see Appendix)
$$
g_L= L^{d-2} \cdot
\frac{4\sqrt{\pi}}{(4\pi)^{d/2}} m^{d-2}
\left(\frac{mL}{2}\right)^{(1-d)/2}  e^{-mL} \,.
\eqno(59)
$$
The origin of the exponential dependence can be
explained in the following manner.  It is well-known \cite{35},
that integration of quickly oscillating functions
$$
f_\omega= \int_{-\infty}^{\infty} f(x)  e^{i\omega x} dx  \,,
\qquad \omega\to \infty
\eqno(60)
$$
involves the analytic properties of $f(x)$. If $f(x)$ has
a jump of the $n$-th derivative at the real axis, then
$f_\omega\sim \omega^{-n-1}$; in particular, the case $n=0$
corresponds usually to integration in finite limits.  If
$f(x)$ is regular at the real axis, then the integration
contour is shifted to the upper half-plane and the integral
is exponentially small,
$$
f_\omega \sim \exp(-const\cdot \omega)  \,.
\eqno(61)
$$
The analogous situation takes place in
approximation of an integral by a discrete sum
$$
\int_{-\infty}^{\infty} f(x)  dx \approx
h \sum_{s=-\infty}^{\infty} \left.f(x_s)\right|_{x_s=hs}\,,
\eqno(62)
$$
which becomes clear after the use of the Poisson summation
formula \cite{36}:
$$
h \sum_{s=-\infty}^{\infty} \left.f(x_s)\right|_{x_s=hs}
= \sum_{k=-\infty}^{\infty}
\int_{-\infty}^{\infty} f(x)\, e^{i2\pi kx/h}\, dx \,.
\eqno(63)
$$
The term with  $k=0$ corresponds to the integral (62), while the
main correction to it has an order of $\exp(-const/h)$. Two sums
over ${\bf q}$ in Eq.58 are equal in the  continual approximation;
their difference is determined by the main effect of discreteness,
which has the order of $\exp(-const\cdot L)$ due to $h\sim 1/L$.

The definition (58) can be written in the form of the
scaling relation
$$
g_L=H_T\left(\frac{L}{\xi_{0D}}\right) \,, \qquad
$$
$$
H_T(z)= \frac{1}{(2\pi)^2} \sum_{\bf t}\,
 \frac{(-1)^{2t_1}} {|{\bf t}|^2+ (z/2\pi)^2 }\,,
\eqno(64)
$$
where we introduced a vector ${\bf t}=(t_1,t_2,\ldots,t_d)$,
whose components $t_i$ run integer values $0,\,\pm 1,\,\pm
2,\,\ldots$ for $i=2,\ldots,d$ and semi-integer values $0,\,\pm
1/2,\,\pm 1$, $\pm 3/2,\,\ldots$ for $i=1$.  The function
$H_T(z)$ is always positive  (Fig.\,8) and has
the asymptotic behavior
$$
H_T(z)= \left \{ \begin{array}{cc} \displaystyle{1/z^2}\,,&
\qquad z\ll 1 \\ (1/\pi) \left( z/2\pi \right)^{(d-3)/2}\,
e^{-z},& \qquad z\gg 1 \end{array} \right.  \,.
\eqno(65)
$$
\begin{figure}
\centerline{\includegraphics[width=2.4 in]{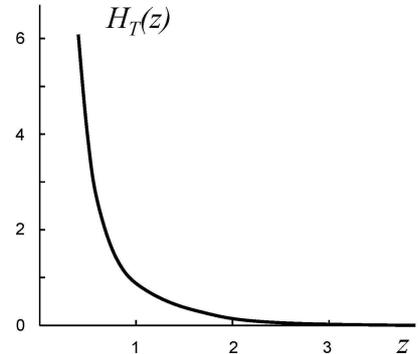}}
\caption{\footnotesize Functions $H_T(z)$,
corresponding to the $d$-dimensional "Thouless definition",
for $d=1,\,2,\,3$: in the scale of the figure all three
curves coincide.} \label{fig8}
\end{figure}

\vspace{4mm}

\begin{center}
{\bf 5. APPLICATION OF THE SHELL MODEL}
\end{center}

The shell model was developed in nuclear physics for
description of coupling between  the discrete spectrum
of the "target" with a continuous spectrum of
scattered particles \cite{210};
Iida et al \cite{18}   suggested to use it for consideration
of the combined system  "sample+external leads".
Below we illustrate this approach using the simple models of solid
state physics and then come to consideration of the many-channel
case.\,\footnote{\,Sec.5 contains derivation of Eq.64 by the
other method and can be omitted by the reader interesting only
in  results. }

\begin{center}
{\bf 5.1. Connection of infinite and finite chains} \end{center}

Consider the model consisting of two chains, upper
infinite ($N\to\infty$) and lower finite (Fig.\,9,a).
We accept for simplicity that the chains are described
by the  usual Anderson model
$$
J\psi_{n+1} + J\psi_{n-1} + \epsilon_n\psi_{n} =E \psi_{n} \,,
\eqno(66)
$$
\begin{figure*}
\centerline{\includegraphics[width=5.1 in]{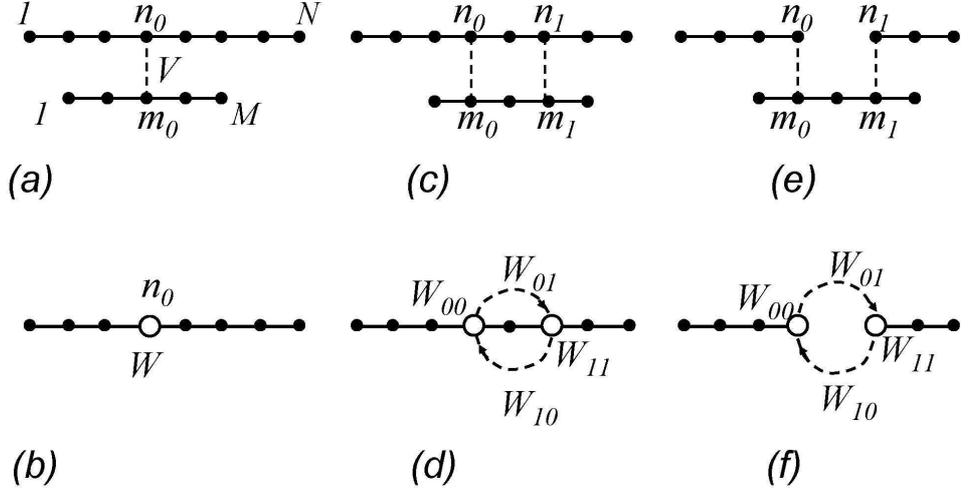}}
\caption{\footnotesize (a) Weak connection  $V$ between two
chains, and (b) the equivalent scheme in the subspace 1.
(c) Two weak connections between chains, and (d) the
corresponding equivalent scheme.  (e) Broken upper chain, and
(f) effective transitions between its parts. }
\label{fig9}
\end{figure*}
though it is irrelevant for the most part of discussion; the upper
chain is supposed to be an ideal conductor ($\epsilon_n= 0$).  As
a perturbation, we include the overlap integral $V$ between the
sites $n_0$ and  $m_0$ of two chains (Fig.\,9,a).  The Hamiltonian
matrix is block-diagonal in the zero approximation, while
perturbation creates non-diagonal elements $V$ in $n_0$-th row and
$m_0$-th column and vice versa; so the matrix elements of the
perturbation operator are
$$
V_{nn'} = V \left( \delta_{nn_0} \delta_{n'm_0} +
              \delta_{n m_0} \delta_{n' n_0}\right)\,.
\eqno(67)
$$
The matrix Dyson equation $G = G_0 + G_0 V G$ written in
components has a form
$$
G_{nn'} = G^0_{nn'}+G^0_{nn_0}V G_{m_0 n'}
                  +G^0_{nm_0}V G_{n_0 n'} \,,
\eqno(68)
$$
where $G_{n n'}$ is the Green function of the perturbed system and
$G^0_{n n'}$ is the initial Green function corresponding to two
independent chains. The index $n$ runs  values  $1,\,2,\,\ldots,N$
corresponding to the first chain, and then values
$N+1,\,N+2,\,\ldots,N+M$ corresponding to the second chain; the
sites of the latter will be also numerated by the index $m$.
Equation (68) is easily solved: setting $n=n_0$ and $n=m_0$ one
obtains the closed system for  $G_{n_0 n'}$ and $G_{m_0 n'}$, and
then its solution is substituted into  (68).  The complete
expression for
 $G_{n n'}$ is rather lengthy, so we give only its
 projection on subspace 1 of the upper chain
 $$
G_{nn'} = G^0_{nn'}+G^0_{nn_0}
  \frac{V^2 G^0_{m_0m_0}}{1 - V^2 G^0_{m_0m_0}G^0_{n_0n_0}}
     G^0_{n_0 n'}
\eqno(69)
$$
and subspace 2 of the lower chain
$$
G_{mm'} = G^0_{mm'}+G^0_{mm_0}
  \frac{V^2 G^0_{n_0n_0}}{1 - V^2 G^0_{m_0m_0}G^0_{n_0n_0}}
     G^0_{m_0 m'}\,.
\eqno(70)
$$
Investigation of (69,\,70) reveals the following
qualitative moments.
\vspace{2mm}

{\it 1. Effective scatterer.} If we are interested only
in movement along the upper chain, then perturbation  (67) is
equivalent to insertion of an impurity atom at the point  $n_0$
(Fig.\,9,b), with the effective Hamiltonian
$$
V_{nn'} = W
\delta_{nn_0} \delta_{n'n_0}\,, \qquad W=V^2 G^0_{m_0m_0}\,.
\eqno(71)
$$
To prove this result,  it sufficient to write down the
Dyson equation for the perturbation (71) and verify that its
solution coincides with  (69).  \vspace{2mm}

{\it 2. Attenuation in the finite system.}
The initial Green function of the lower chain has
a form
$$
G^0_{mm'} = \sum_{s} \frac{e_s(m)e_s^*(m')}{E-\epsilon_s+i0} \,,
\eqno(72)
$$
where  $\epsilon_s$ and  $ e_s(m)$ are its eigenvalues and
eigenvectors. In the vicinity of a level  $\epsilon_s$ the sum is
determined by one term; its substitution to (70) gives
$$
G_{mm'} = \frac{e_s(m)e_s^*(m')}
{E-\epsilon_s - V^2 G^0_{n_0n_0}|e_s(m_0)|^2 }\,
.
\eqno(73)
$$
For the ideal chain   $G^0_{nn}$ does not depend  on $n$,
and
$$
G^0_{nn} = \int \frac{dk}{2\pi}
\frac{1}{E-\epsilon(k)+i0} \equiv I(E)-i\pi\nu(E)\,,
\eqno(74)
$$
where $\nu(E)$ is the density of states at the energy $E$.
Since (81) is valid for any level $\epsilon_s$, and for small
$V$ we can neglect the mutual influence of different levels,
then the effective Green function of the lower chain can be
written as
$$
\tilde G_{mm'} = \sum_s \frac{e_s(m)e_s^*(m')}
{E-\tilde\epsilon_s +i\gamma_s }\,,
\eqno(75)
$$
i.e. the discrete levels acquire a finite attenuation
$$
\gamma_s=\pi V^2 \nu(\epsilon_s) |e_s(m_0)|^2
\,.\eqno(76)
$$
The difference between $\tilde\epsilon_s$ and $\epsilon_s$ has no
qualitative effect and can be neglected for small $V$.

\vspace{2mm}
{\it 3. Effective  $T$ matrix.} The combination
entering (69),
 $$
 T=   \frac{V^2 G^0_{m_0m_0}}{1 - V^2
G^0_{m_0m_0}G^0_{n_0n_0}}  \,,
\eqno(77)
$$
is in fact the  $T$ matrix of  scattering \cite{37};
by definition, its substitution into the Born expression instead
of the perturbation $V$ gives the exact scattering
amplitude. Considering it in the vicinity of a level
$\epsilon_s$, one finds  possibility to write it
in the form
$$
T \approx V^2 \tilde G_{m_0m_0} \,.
\eqno(78)
$$
It differs from the Born result  $T =V^2 G^0_{m_0m_0}$
(following from the concept of the effective scatterer)
by replacement of  $G^0$ by $\tilde G$, i.e. by taking
attenuation of states into account.


\begin{center}
{\bf 5.2. Several bonds between chains} \end{center}

The simplest generalization
of the model contains several bonds between chains,
connecting the pairs of sites  $n_0$ and $m_0$, $n_1$ and
$m_1$, and so on  (Fig.\,9,c). In this case the perturbation
operator is defined as
$$
V_{nn'} = V \sum_i \left( \delta_{nn_i} \delta_{n'm_i} +
            \delta_{n m_i} \delta_{n' n_i}\right) \,,
\eqno(79)
$$
and the Dyson equation reads
$$
G_{nn'} = G^0_{nn'}+ \sum_i \left(G^0_{nn_i}V G_{m_i n'}
                  +G^0_{nm_i}V G_{n_i n'}\right) \,.
\eqno(80)
$$
Let  $n$ and $n'$ belong to the upper chain; then
$G^0_{nm_i}=0$ and  (80) accepts the form
$$
G_{nn'} = G^0_{nn'}+ \sum_i G^0_{nn_i}V G_{m_i n'}
\,.
\eqno(81)
$$
On the other hand, setting $n=m_i$ in (80)
$$
G_{m_in'} =  \sum_j G^0_{m_im_j}V G_{n_j n'}
\eqno(82)
$$
and substituting in (81), we obtain
$$
G_{nn'} = G^0_{nn'}+ \sum_{ij} G^0_{nn_i}
\cdot V^2 G^0_{m_i m_j} \cdot
G_{n_j n'} \,.
\eqno(83)
$$
If only the upper chain is of interest, then the
effective perturbation Hamiltonian
$$
V_{nn'} = \sum_{ij} W_{ij} \delta_{nn_i} \delta_{n'n_j}\,,
\qquad W_{ij}=V^2 G^0_{m_im_j}\,
\eqno(84)
$$
can be used, i.e. the scatterers  $W_{ii}$ are introduced
in the points $n_i$ and the additional overlap integrals
$W_{ij}$ are included
between points  $n_i$ and $n_j$ (Fig.\,9,d).

\begin{center}
{\bf 5.3. Broken upper chain}
\end{center}

Let us
remove the portion between  $n_0$ and $n_1$ in the upper
chain (Fig.\,9,e). The above expressions formally retain,
 if  the initial Green  function $G^0_{nn'}$ is taken
 for  the broken chain. The equivalent scheme shows
 (Fig.\,9,f), that transitions between
the left and right parts of the upper chain
 are
 possible only due to overlap integrals $W_{01}$ and $W_{10}$,
 so the transmission coefficient is proportional to $|W_{10}|^2$.

\begin{center}
{\bf 5.4. Many-channel case}
\end{center}

Now introduce the model which has the immediate interest
(Fig.\,10,a): the lower chain is replaced by a finite
$d$-dimensional system, whose points $m_0^{(i)}$ and $m_1^{(i)}$
are related with sites  $n_0^{(i)}$ and $n_1^{(i)}$ of
the ideal one-dimensional chains.
This model describes a situation, when
the given $d$-dimensional system
is weakly connected with the ideal leads.
\begin{figure*}
\centerline{\includegraphics[width=5.1 in]{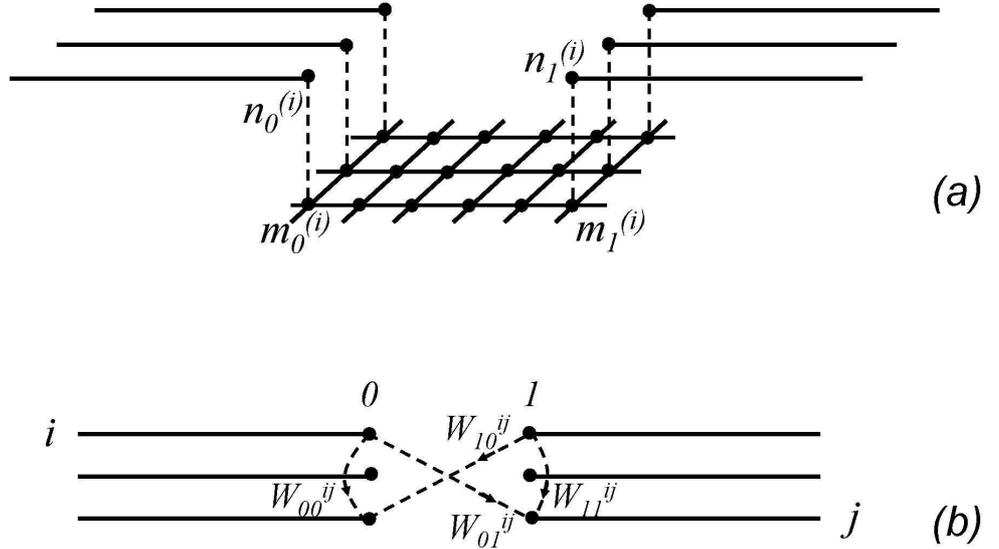}}
\caption{\footnotesize (a) The modelling of the system
"sample\,+\,external leads". Ideal one-dimensional chains are
weakly connected with the considered $d$-dimensional
system.  (b) Effective transitions between chains
corresponding to many-channel scattering matrix
(see Fig.2,c).} \label{fig10}
\end{figure*}

Expressions  (79\,--\,84) are formally applicable, if the
index $n_{i}$ runs  values  $n_0^{(i)}$ and $n_1^{(i)}$,
while the index $m_{i}$ runs values $m_0^{(i)}$ and
$m_1^{(i)}$.  Therefore, we can immediately introduce the
equivalent scheme (Fig.\,10,b), according to which the $i$-th lead
on left and the  $j$-th lead on right are related by overlap
integrals $W_{00}^{ij}$, $W_{01}^{ij}$, $W_{10}^{ij}$,
 $W_{11}^{ij}$.  Comparing with the Landauer many-channel
 scattering matrix (Fig.\,2,c), we see that the amplitude  $t_{js}$
 are determined by the quantities $W_{10}^{sj}$,
$$
t_{js} = -2i e^{2ik_F}\sin{k_F} \frac{ W_{10}^{sj}}{J} \,,
\eqno(85)
$$
where  $k_F$ is the Fermi momentum.
To derive this relation, one should write the effective
Schroedinger equation
$$
E \psi_{n}= J\psi_{n+1} + J\psi_{n-1} + \epsilon_n\psi_{n}+
$$
$$
    +\sum_{sj}\left[
           W_{00}^{sj} \delta_{nn_0^{(s)}}\psi_{n_0^{(j)}}
          +W_{01}^{sj} \delta_{nn_0^{(s)}}\psi_{n_1^{(j)}}
                                   \right.
$$
$$
\left.    +W_{10}^{sj} \delta_{nn_1^{(s)}}\psi_{n_0^{(j)}}
          +W_{11}^{sj} \delta_{nn_1^{(s)}}\psi_{n_1^{(j)}}
                                   \right]
\eqno(86)
$$
and find the solution of the scattering problem: if a
 wave of the unit amplitude is incident to the channel  $s$,
then the amplitude in  $j$-th channel can be written
in the form  ($k$ is a wavenumber)
$$
\psi_n^{(j)}=\left \{ \begin{array}{cc}
\delta_{sj}e^{ik(n-n_0)}+r_{sj}e^{-ik(n-n_0)}\,,& n\le n_0 \\
t_{sj}e^{ik(n-n_1)}\,, & n\ge n_1 \end{array} \right.\,.
\eqno(87)
$$
It leads to  the system of equations
$$
J e^{ik}\delta_{sj} + J e^{-ik} r_{sj} = W_{00}^{js} +
$$
$$
+\sum_{j'} \left( W_{00}^{jj'} r_{sj'}+
                     W_{01}^{jj'} t_{sj'} \right)
$$
$$
J e^{-ik}t_{sj}  = W_{10}^{js}
+\sum_{j'} \left( W_{10}^{jj'} r_{sj'}+
                     W_{11}^{jj'} t_{sj'} \right) \,,
\eqno(88)
$$
whose iterations in $W_{\alpha\beta}^{js}$ give (85).
Substituting (85) in the Landauer formula (7),
one has
$$
G_L = \frac{e^2}{2\pi\hbar}\, 4\sin^2{k_F}\, \frac{V^4}{J^2}
\,\sum_{ij} \left| G_{m_0^{(i)} m_1^{(j)}}^{0} \right|^2
\eqno(89)
$$
 This result corresponds to
the Born approximation: to obtain the complete result, one
should find the many-channel  $T$ matrix.

\begin{center}
{\bf 5.5. $T$ matrix in the many-channel case} \end{center}

If $\Phi(r)=e^{i\rm k\cdot r}$ is a plane wave and
$\Psi(r)$ is a solution of the scattering problem, then they
are related by the Lippmann--Schwinger equation \cite{37}
$$
\left| \Psi\right> = \left| \Phi\right> +G_0 V
\left| \Psi\right> \,,
\eqno(90)
$$
which can be iterated as
$$
\left| \Psi\right> = \left\{ 1+G_0 V+G_0 V G_0 V+
\right.
$$
$$
\left.
+G_0 V G_0 V G_0 V + \ldots\right\}
 \left| \Phi\right>  \,.
\eqno(91)
$$
Let us set  $G_0=G_1+G_2$, where  $G_2$ corresponds to
the system under consideration, and  $G_1$ to ideal leads;
then perturbation  $V$ relates only $G_1$ and  $G_2$, while
combinations  $G_1VG_1$ and $G_2VG_2$ turn to zero.
Accepting that states  $\left|
\Psi\right>$ and $ \left|\Phi\right>$ belong to subspace 1,
we have
$$
\left| \Psi\right> = \left\{ 1+G_1 V G_2 V+ \right.
$$
$$
+G_1 V G_2 V  G_1 V G_2 V +
\left.\ldots\right\}
 \left| \Phi\right>  \,.
\eqno(92)
$$
By definition,  $\left| \Psi\right>$ and $ \left|
\Phi\right>$ are related by the $S$ matrix, $\left|
\Psi\right>=S \left| \Phi\right>$, and (92) gives
$$
S=1+G_1 V G_2 V\frac{1}{1-G_1 V G_2 V}  \,.
\eqno(93)
$$
The $T$ matrix is introduced by the relation
$V\left| \Psi\right>=T\left| \Phi\right>$ \cite{37},
so $S=1+G_0 T$,  reducing to $S=1+G_1 T$
in the subspace 1; so
$$
T= V G_2 V\frac{1}{1-G_1 V G_2 V}=
$$
$$
=V \frac{1}{E-H_2-V G_1 V} V  \,,
\eqno(94)
$$
where the relation $G_2=(E-H_2)^{-1}$ is used. The poles of
the $T$ matrix are determined by eigenvalues of the operator
$H_2+V G_1 V$, which can be found perturbatively
$$
\lambda_s =\epsilon_s+
\left< e_s\right|V G_1 V  \left| e_s\right>  \,.
\eqno(95)
$$
Using the specific form (79) of the matrix elements of $V$,
we have  $\lambda_s =\tilde\epsilon_s -i\gamma_s $,
where  $\gamma_s$ are determined by expression
$$
\gamma_s=\pi V^2 \nu_F \sum_i |e_s(m_i)|^2
\,,\eqno(96)
$$
which is a natural generalization of (76). This result can be
also obtained by induction, including connections one after
another and neglecting their influence  on $G_{nn'}^{0}$.

Below we are interested in the limit of small $V$. In this case
we can take into account only the
qualitative effect related with attenuation,
neglecting influence of perturbation on
eigenfunctions and eigenvalues. In the Born
approximation, the  $T$ matrix has a form  $VG_2V$  (see
(94)) and has the poles at $\epsilon_s$. Substitution of
$\epsilon_s-i\gamma_s$ for $\epsilon_s$  corresponds to
replacement of  $G^{0}$ by  $\tilde G$ in expression (89)
$$
G_L = \frac{e^2}{2\pi\hbar}\, 4\sin^2{k_F}\, \frac{V^4}{J^2}
\,\sum_{r_{\bot},r'_{\bot}} \left| \tilde G({\bf r},{\bf r'})
\right|^2_{|x'-x|=L}
\eqno(97)
$$
where  $\tilde G$ is defined analogously to (75). Here we
introduced the longitudinal ($x$) and transverse ($r_{\bot}$)
components of vector  $r$, and summation occurs over the points of
connection with leads. Taking the average value of the
conductance, we can consider $\langle|\tilde G|^2\rangle$ as a
zero-frequency limit of the density correlator (49) and use it in
the form  (52). Consequently, we have related the Landauer
conductance with the diffusion coefficient $D(\omega,q)$. Since
(97) corresponds to the open system, the replacement  (53) is
implied, leading to a finiteness of the diffusion coefficient
$D_L$.  Omitting (here and later) irrelevant constant factors we
have for the dimensionless conductance
$$
g_L =  \frac{V^4}{J^4} \frac{J}{D_L}
\,\sum_{r_{\bot},r'_{\bot}} \left.
K({\bf r},{\bf r'})\right|_{|x-x'|=L}   \,,\qquad
$$
$$
K({\bf r},{\bf r'})= \frac{1}{L^d}
\sum\limits_{\rm q}\,\frac{e^{i\rm q\cdot(r-r')}}{m^2 +
q^2} \,.
\eqno(98)
$$

\begin{center}
{\bf 5.6. Conductance of a finite system: definition}
\end{center}

Let a finite system has a form of the $d$-dimensional
cube connected to external leads, composed of
$N_c$ ideal one-dimensional chains; one should differ the
"thin" and "bulk" contacts (Fig.\,11). In the first case all
\begin{figure}
\centerline{\includegraphics[width=2.5 in]{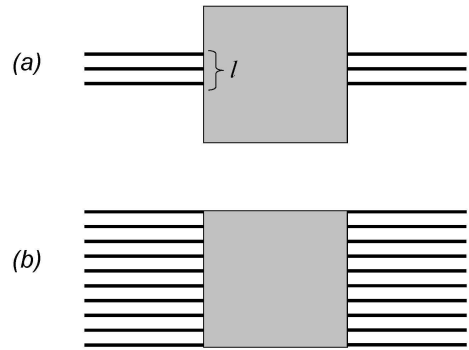}}
\caption{\footnotesize The  "thin" (a) and  "bulk" (b)
contacts attached to the system.} \label{fig11}
\end{figure}
chains are connected to the spot of a size $l\ll \sqrt{L\xi}$
(where  $\sqrt{L\xi}$ is a characteristic scale where
$K(r,r')$ essentially changes as a function of the
transverse coordinate $r_{\bot}$), in the second case
they are uniformly distributed along the side of the cube.
For  the "thin" contacts we can set  $r_{\bot}=0$ and write
$$
g_L =  \frac{V^4}{J^4} N_c^2 \frac{J}{D_L}
\,\left.  K(x,x')\right|_{|x-x'|=L}
\eqno(99)
$$
In the metallic state  $|e_s(m)|^2\sim L^{-d}$ and according
to (96) attenuation of all states has the same order of
magnitude
$$
\gamma \sim V^2 \nu_F N_c L^{-d} \sim
\frac{V^2}{J^2} \,N_c \,\Delta \,,
\eqno(100)
$$
where $\Delta$ is the level spacing. It is convenient to
introduce the parameter
$$
k_b= \frac{V^2}{J^2} \,N_c \,,
\eqno(101)
$$
having a sense of the effective transparency of an interface.
According to (99), $g_L$ contains a factor  $k_b^2$ in
the explicit form and dependence
$D_L\propto \gamma \propto k_b$ in the
diffusion constant, so $g_L\propto k_b$. The
 proportionality coefficient  can be estimated from
the condition that for  $k_b\sim 1$ attenuation  $\gamma$ is
of the order $\Delta$ and according to scaling theory
(see (1,2)) the block of size $L$ is in the critical
regime, i.e.  $g_L\sim 1$ and $D_L\sim JL^{2-d}$:
$$
g_L = k_b L^{d-2} \,\left.  K(x,x')\right|_{|x-x'|=L} \,.
\eqno(102)
$$
This result is valid for $k_b\alt 1$, when  perturbation
theory is applicable. In the region $k_b\agt 1$,
one expects the absence of the $k_b$ dependence, since
$\gamma\agt \Delta$ and
the extended levels overlap strongly and
form the practically constant
density of states (Fig.\,4).  It is easy to see (Fig.\,12), that
\begin{figure}
\centerline{\includegraphics[width=2.5 in]{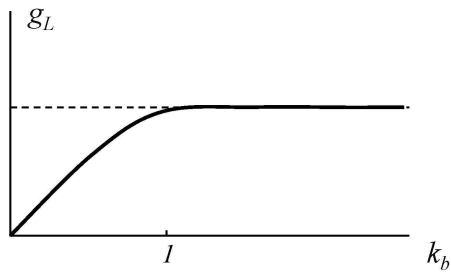}}
\caption{\footnotesize Conductance $g_L$ of a
finite system versus degree of its openness.}
\label{fig12}
\end{figure}
the conductance of the maximally open system is obtained from (102) at
$k_b\sim 1$.\,\footnote{\,Physically, the most adequate estimate
of $g_L$ for the extremely open system corresponds to the plateau
at $k_b\agt 1$ in Fig.\,12; however, it is not reasonable to use
very large values of  $k_b$, since the  "plateau" can in fact be a
slow $k_b$ dependence due to influence of environment on the
system. Note, that numerical modelling  \cite{21} usually deals
with the limit  $k_b\to\infty$.}
However, the factor  $k_b$ can be eliminated from (102) not
only setting  $k_b\sim 1$, but also taking the derivative
at $k_b\to 0$:
$$
g_L^{open} = \left. \frac{d g_L(k_b)}{d k_b}\right|_{k_b=0}
           = L^{d-2} \,\left.  K(x,x')\right|_{|x-x'|=L} \,.
\eqno(103)
$$
We accept (103) as a definition for the conductance of
a finite system: physically, it corresponds to the
extremely open system, but is formulated in terms of
almost closed systems.
Due to the latter, such definition reflects the internal
properties of the given system, not disturbed by its environment.
Simultaneously, it provides the elegant solution of the contact
resistance problem (Sec.\,1): in the small $k_b$ limit one can use
the two-probe formulas (5,7) of Economou--Soukoulis type  (due to
$t_{ij}\to 0$) and there is no need in the original Landauer
formula (4) or its ambiguous multi-channel generalizations
\cite{3,8,11,13}.
The allowed values of ${\bf q}$
in the sum (98) for  $K({\bf r},{\bf r'})$
 correspond to the closed system
and include the value  ${\bf q}=0$, so  $g_L$
diverges at $m\to 0$ and the conductance of the ideal system
($\xi_{0D}=\infty$) appears to be infinite. It brightens
one  of the  widely discussed questions  \cite{15}.

In the localized phase,
transition from  (99) to (103)  requires more
complicated argumentation. The estimate (100) for $\gamma$ retains
for  $L\alt \xi$; according to it,  condition  $k_b\sim 1$
corresponds to the critical regime for the block of size  $\xi$.
Let $k_b\gg 1$; then the block of size  $\xi$ is in the metallic
state, and hence the layer of width  $\sim \xi$ around contacts is
metallized (Fig.\,13,a). Let us approximate this metallic region
\begin{figure}
\centerline{\includegraphics[width=2.7 in]{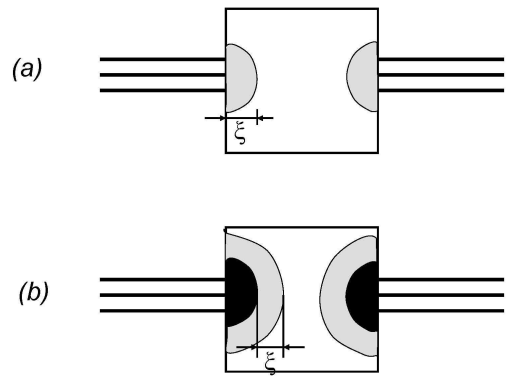}}
\caption{\footnotesize A situation in the localized phase. The
contacts attached to the system arouse metallization of
a layer of width  $\sim \xi$ around them (a). If the
metallized region is replaced by the ideal conductor,
then the contacts shift into the deep of the system and
arouse metallization of the next layer of width
$\sim \xi$ (b).} \label{fig13}
\end{figure}
by the ideal conductor: then contacts shift into the deep of the
system and arouse metallization of the next layer of the width
$\sim \xi$ (Fig.\,13,b), and so on. It is easy to see that a
condition  $\gamma\gg\Delta$ is valid at all length scales till
size $L$. This picture of successive metallization is valid for
$k_b\gg 1$, but retains marginally at  $k_b\sim 1$: in this case
the blocks of size $\xi$ are in the critical regime and the law of
their composition reduces to stationarity of $g_L$, or $D_L\sim J
L^{2-d}$, as in the metallic phase. For $k_b\ll 1$ metallization does
not occur and the condition $\gamma\ll\Delta$  is valid at all
length scales. One can see that the estimate  (102) and the
$k_b$ dependence   (Fig.\,12) remain the same as in  the metallic
phase.

\begin{center}
{\bf 5.7. Equivalence with  the "Thouless definition"}
\end{center}

The allowed values  of  ${\bf q}$ in the sum  (98) correspond to
the closed system. Assuming the latter to be a system with
periodical boundary conditions, it is natural to accept its size
to be $L$ in the transverse direction and $2L$ in the longitudinal
direction:  then for $|x-x'|=L$ the contacts are arranged at the
opposite sides of the cylinder, in which the system is effectively
coiled.  Considering the 1D case for simplicity, we have
$$
K(x,x')= \frac{1}{2L} \sum\limits_s
\,\left.\frac{e^{i q_s L}}
{q_s^2+ m^2 }\right|_{q_s=2\pi s/2L}\,
\eqno(104)
$$
and noticing that $e^{i q_s L}=(-1)^s$ we can
separate the terms with odd and even  $s$,
$$
K(x,x')= \frac{1}{2L} \left(
 \sum\limits_s\,\left.\frac{1}
{q_s^2+ m^2 }\right|_{q_s=2\pi s/L}\,-
\right.
$$
$$
\left. -
  \sum\limits_s\,\left.\frac{1}
{q_s^2+ m^2 }\right|_{q_s=(2\pi s+\pi)/L}
           \right)\,.
\eqno(105)
$$
For the Bloch boundary conditions (47), the allowed values are
$q_s=(2\pi s +\varphi)/L$ and Eq.105 contains the
difference of the terms with $\varphi=0$ and $\varphi=\pi$,
so (103) is equivalent to the "Thouless definition"  (58).

For the more realistic boundary conditions (48),
the eigenfunctions of the operator $-\partial^2/\partial x^2$
has a form  $A_s\sin(q_sx+\psi_s)$ with
$$
A_s^2=\frac{2}{L+2\kappa/(q_s^2+\kappa^2)}\,,
\qquad \psi_s=\arctan(q_s/\kappa)   \,,
\eqno(106)
$$
and the allowed values of $q_s$ are determined by equation
$$
q_sL+2\arctan(q_s/\kappa)=\pi s\,, \qquad s=1,2,3,\ldots
\eqno(107)
$$
The expression for $K(x,x')$ has a form
$$
K(x,x')= \sum\limits_{s=1}^\infty\,
A_s^2 \,\frac{\sin(q_sx+\psi_s)\sin(q_sx'+\psi_s) }
{q_s^2+ m^2 }\,,
\eqno(108)
$$
and for the closed system ($\kappa\to 0$) reduces to
$$
K(0,L)= \frac{1}{L}\sum\limits_{s=-\infty}^\infty\,
\left.\frac{\cos(q_sL)}
{q_s^2+ m^2 }\right|_{q_s=\pi s/L}\,.
\eqno(109)
$$
We have transformed the product of sines into the difference
of cosines and extended the summation to negative  $s$, using
evenness in  $q_s$.
Noting that $\cos{q_sL}=(-1)^s$, one can separate odd and even
$s$ and obtain the result coinciding with  (105) apart from the
irrelevant constant factor. The accepted limitation by one
dimension is not essential: the $d$-dimensional case differs only
by summation over transverse components of  ${\bf q}$, which is
the same for two terms of the difference (105).

Below we had in mind the case of the "thin" contacts (Fig.\,11,a).
For the "bulk" contacts (Fig.\,11,b) we have instead (103)
$$
g_L^{open} =  L^{d-2} \,\frac{1}{N_c^2}
\,\sum_{r_{\bot},r'_{\bot}} \left.
K({\bf r},{\bf r'})\right|_{|x-x'|=L} \,.
\eqno(110)
$$
If the one-dimensional chains are connected to each
site on the plane of the cube, then  $N_c=L^{d-1}$ and summation
over $r_{\bot},r'_{\bot}$ removes the transverse
components of the vector ${\bf q}$, so (110) reduces to the
result for the 1D case. We see that the natural definitions
for the conductance of a finite system are exhausted by
$d$-dimensional "Thouless definitions". The intrinsic
$d$-dimensional case is realized for the "thin" contacts
(Fig.\,11,a).  The effective dimensionality is diminished by the
unity if one-dimensional chains are connected along the line,
which goes through the whole plane of the cube. For the "bulk"
contacts (Fig.\,11,b) the effective dimensionality is
unity.

We should notice, that the physical considerations define
$g_L$ to the factor of the order of unity. Such
uncertainty is natural and related with an arbitrary choice
of the unit scale. Only ratios of conductances are relevant,
while the choice of the absolute scale if a subject of
convention.

We see that one of two scaling relations (10,\,12)
(written as (66) in Sec.4) can be obtain from the pure
quantum mechanical consideration without use of the
self-consistent theory of localization. The second scaling
relation can be also studied by other methods: in this case,
the quantity $\xi_{0D}$  should be defined by Eq.11,
where $D(\omega,0)$ is the diffusion coefficient
of the closed system.

\begin{center}
{\bf 6. DISCUSSION OF SCALING EQUATIONS} \end{center}

According to Secs.\,3,\,4, dependence of  $g_L$ on
$L/\xi$ is represented in the parametric form
$$
\pm c_d \left(\frac{L}{\xi} \right)^{d-2}
= H \left(z \right) \,,\qquad g_L=H_T(z)
\eqno(111)
$$
for  $2<d<4$, and
$$
\frac{1}{2\pi}\ln\left(\frac{\xi}{L} \right)
= H \left(z \right)  \,,  \qquad g_L=H_T(z)
\eqno(112)
$$
for  $d=2$; in $d<2$ dimensions, representation  (111) holds
with the upper sign and consists of one branch. Using
asymptotic behavior of  $H(z)$ (33,\,38,\,40) and
$H_T(z)$ (65), one obtains for the length dependence of
$g_L$  (Fig.\,14) at $d>2$:
\begin{figure}
\centerline{\includegraphics[width=2.5 in]{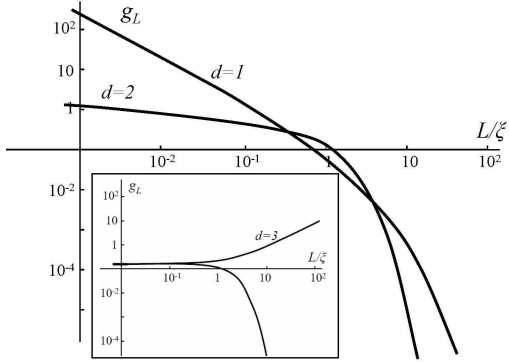}}
\caption{\footnotesize Conductance  $g_L$ versus  $L/\xi$
for  $d=1,\,2,\,3$.} \label{fig14}
\end{figure}
$$
g_L=
\left \{ \begin{array}{cc}
\displaystyle{c_d(L/\xi)^{d-2}}\,,& \, g_L\gg 1 \\
\displaystyle{g_c\pm B(L/\xi)^{d-2}}\,,& \, g_L\to g_c \\
(1/\pi) \left( L/2\pi\xi \right)^{(d-3)/2}\, e^{-L/\xi},&
\, g_L\ll 1 \end{array} \right.  \,,
\eqno(113)
$$
where  $g_c=H_T(z^*)$, $B=c_d H'_T(z^*)/H'(z^*)$. For $d< 2$ the
results for small and large  $g_L$  are formally the same, but the
critical point  $g_c$ is absent. For $d=2$, the result in the
metallic phase $g_L=(1/2\pi)\ln(\xi/L)$ can be represented as a
logarithmic correction to the Drude conductance $g_0$,
$g_L=g_{0}-(1/2\pi)\ln(L/a)$ \cite{1}, with the asymptotics (113)
for  $g_L\ll 1$.

The Gell-Mann -- Low function $\beta(g)$ is determined by
the derivative $d\ln g/d\ln L$  (see Eq.2) and can be written
in the parametric form
$$
 g=H_T(z)\,,\qquad \beta(g)=(d-2)\frac{H(z)H'_T(z)}
 {H_T(z)H'(z)}
\eqno(114)
$$
for  $d\ne 2$, and
$$
 g=H_T(z)\,,\qquad \beta(g)=-\frac{1}{2\pi}\frac{H'_T(z)}
 {H'(z) H_T(z)}
\eqno(115)
$$
for $d=2$. Since  $H_T(z)$ is positive, while
$H'_T(z)$ and $H'(z)$ are negative (Figs.6,\,8),
the  $\beta$ function is negative for $d=2$, and has a root
for $d>2$ due to the root of  $H(z)$. The calculated
 $\beta$ functions for  $d=1,\,2,\,3$ are shown in Fig.\,5.

The expansions of  $H(z)$ and  $H_T(z)$ in powers of  $z^2$
has a following form for  $d<2$
$$
H(z)=1/z^2+a_0-a_2 z^2 +a_4 z^4-a_6 z^6+\ldots\,,
$$
$$
H_T(z)=1/z^2+\tilde a_0-\tilde a_2 z^2 +\tilde a_4 z^4
-\tilde a_6 z^6+\ldots\,,
\eqno(116)
$$
where
$$
 a_{2n}=  \sum_{{\bf s}\ne 0}\,
 \frac{1} {(2\pi|{\bf s}|)^{2n+2}}\,, \qquad
\tilde a_{2n}=  \sum_{{\bf t}\ne 0}\,
 \frac{(-1)^{2t_1}} {(2\pi|{\bf t}|)^{2n+2}}\,,
\eqno(117)
$$
and vectors  ${\bf s}$ and ${\bf t}$ are the same as in
(28) and (64).  In the case  $d\ge 2$, the coefficient
$a_0$ is replaced by the parameter  $b_1$, introduced in
(29) and (36). In the case  $d=1$, the coefficients $a_{2n}$
and  $\tilde a_{2n}$ are expressed in terms of the Riemann
$\zeta$ function or the Bernoulli numbers \cite{36} and can be
obtained from the closed expressions
$$
H(z) =  \,\frac{1}{2z \,\tanh(z/2)} \,, \qquad
H_T(z) =  \,\frac{1}{z \,\sinh{z}} \,,
\eqno(118)
$$
following from  (39) and (64) with the use of the Poisson
summation formula.\,\footnote{\,These results can be used for
numerical calculations in higher dimensions, in order to produce
the analytic summation along one of coordinate axes. }
Using (116), one can find the expansion of  $\beta(g)$
in powers of  $1/g$
$$
\beta(g)=(d-2)+\frac{(d\!-\!2)(a_0\!-\!\tilde a_0)}{g}+\ldots\,,
\quad d\ne 2
\eqno(119)
$$
$$
\beta(g)=-\frac{1}{2\pi g}+\frac{a_2\!-\!\tilde
a_2}{2\pi g^3}+\ldots\,,\quad d=2 \,.
\eqno(120)
$$
The latter result can be compared with the
expansion obtained in the $\sigma$ model approach
\cite{31,32}
$$
\tilde\beta(t)=
-2t^2+0\cdot t^3
+0\cdot t^4 - 12\zeta(3) t^5
+\ldots\,,
\eqno(121)
$$
which is written in terms of the variable  $t\sim 1/g$.
Recalculating  (120) to the same form, one has
$$
\tilde\beta(t)= -2t^2+0\cdot t^3 +
32\pi^2(a_2\!-\!\tilde a_2) t^4 +\ldots\,.
\eqno(122)
$$
The first two coefficients coincide with  (121), while the third
one depends on details of the $g_L$ definition, since the
parameter  $\tilde a_2$ is different for the  "thin" and "bulk"
contacts (Fig.11).  Such situation is well-known in the quantum
field theory  \cite{38,39,40}, where the structure of
expansion for the $\beta$ function is the same as (121,\,122): the
first two coefficients are invariant, while the rest depend on the
renormalization scheme. Transformation from one scheme to another
corresponds to the change of variables  $\tilde t=f(t)$, relating
two different definitions of the charge; expansion of  $f(t)$ in
the series and the proper choice of the coefficients allows to
transform  (122) into  (121) \cite{39,40}. The function  $f(t)$ is
well-defined in perturbation theory but can be singular at $t\sim
1$, indicating that one of two schemes is surely defective
\cite{40}.  Since (122) corresponds to the physical definition of
$g_L$,  such problems can refer only to the expansion  (121). In
the framework of perturbation theory Eqs.121 and  122 are
completely equivalent.

Such equivalence is destroyed in the space dimension
$d=2+\epsilon$. The dimensional regularization used in
the $\sigma$ models corresponds to the $\beta$ function of
the form
$$
\beta_{2+\epsilon}(g)= \epsilon +\beta_{2}(g) \,,
\eqno(123)
$$
i.e.  the $d$ dependence  is present only in the first term
of $1/g$ expansion. The exponent  $\nu$ is determined by the
derivative of $\beta(g)$ at the fixed point, and the
corresponding result \cite{31}
$$
\nu =1/\epsilon -(9/4) \zeta(3) \epsilon^2+\dots\,
\eqno(124)
$$
is in conflict  with (16). This fact is  usually
considered as a proof that the Vollhardt and W$\ddot o$lfle
theory cannot be exact.

However, another interpretation is possible. Let us assume
 (in accordance with  \cite{41,33}), that the
Vollhardt and W$\ddot o$lfle theory is correct: then the
physical reality consists in
existence of the exact result
$\nu=1/\epsilon$ for  $d=2+\epsilon$ and the non-trivial
$\beta$ function for  $d=2$.  The latter  is  related
with the physical essence of the problem:
the logarithmic behavior at $g\ll 1$ (see Eq.3)
makes impossible for $1/g$ expansion to be truncated
 at finite number of terms.
Such physical reality is incompatible with
the formalism of  dimensional regularization:  according to
(123), the exact result  $\nu=1/\epsilon$ is possible only for
the trivial function  $\beta_2(g)=A/g$. Description of reality
with such formalism should lead to unsolvable problems.
Exactly such situation takes place in the modern theory:
the Anderson transition problem reduces to the $\sigma$ model in
a certain approximation but the corresponding renormalization
group is unstable to high gradient terms
\cite{42,43}. It is interesting to
to carry out  renormalization of
$\sigma$ models without the use of  dimensional
regularization: there are indications that in this case the
high-gradient catastrophe is absent  (see discussion
of \cite{44} in the paper \cite{43}).

The latter asymptotics in (113) can be compared
with the exact results for the 1D case \cite{45}
$$
\langle g \rangle= {\textstyle \frac{\pi}{2}}
\left( \alpha L/\pi \right)^{-3/2}\, e^{-\alpha L/4}  \,,
$$
$$
\exp\langle\ln g \rangle= 4\,
e^{-\alpha L}
\,,
\eqno(125)
$$
$$
\left\langle 1/g \right\rangle={\textstyle \frac{1}{2}}
 \, e^{2\alpha L}  \,,
$$
where  $\alpha\propto W^2$ for weak disorder.
The effective correlation length in the dependence $\exp(-L/\xi)$
is sensitive to details of the averaging procedure  and
determined only by the order-of-magnitude, in correspondence
with its physical sense.  The same uncertainty exists in the
present theory, where $\xi$  is defined by  the relation
$D(\omega)=(-i\omega)\xi^2$ with the ambiguous choice of the
absolute scale for $D(\omega)$. Within such
uncertainty, there is no sense to discuss the
precise form of the pre-exponential dependence which corresponds
to redefinition of $\xi$ by the factor  $1+O(\ln L/L)$.

\begin{center}
{\bf 7. COMPARISON WITH NUMERICAL AND PHYSICAL
EXPERIMENTS } \end{center}

Behavior of $g_L$ versus  $L/\xi$  for  $d=3$ is
compared (Fig.\,15,a) with numerical results by Zharekeshev
\cite{30}, where  $g_L$ was estimated as
\begin{figure*}
\centerline{\includegraphics[width=5.1 in]{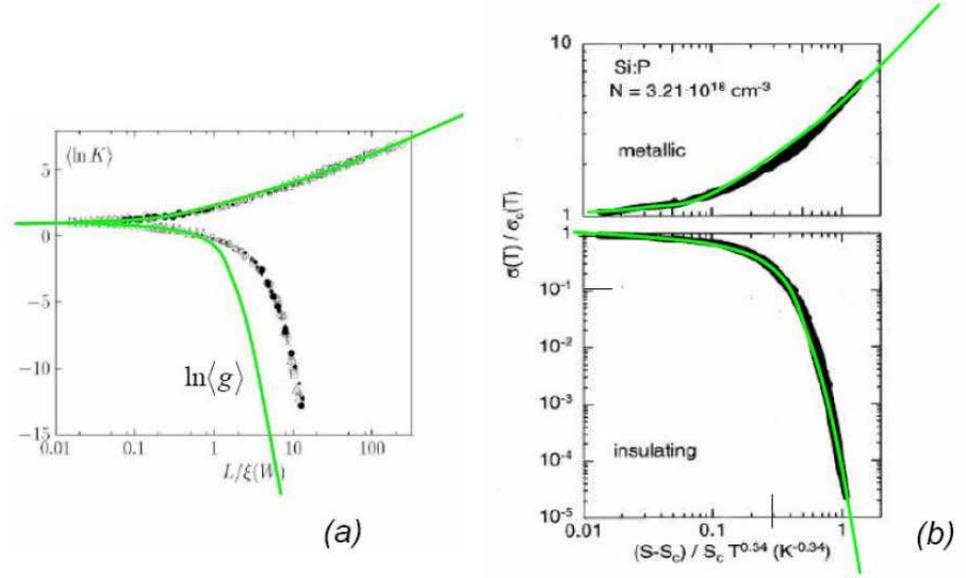}}
\caption{\footnotesize (a) Comparison of theoretical scaling
curves for $d=3$ with numerical results by Zharekeshev
\cite{30}. The general form of the curves was determined without
adjustable parameters, and only parallel shifts along two axes
were made (the same for two branches), corresponding to
the choice of the absolute scales for  $g_L$ and $\xi$.
(b) Comparison of the same curves with the experimental
results for Si-P  \cite{49} under assumption
$L\propto T^{-\alpha}$, where  $\alpha$ was chosen
independently for two branches.} \label{fig15}
\end{figure*}
"acceleration" of levels  $K_s=d^2 \epsilon_s/d\varphi^2$ at
$\varphi=0$
($\varphi$ is the parameter in the
boundary condition (47)), which is
a possible  variant of the Thouless definition \cite{17}.
The agreement is satisfactory in the metallic state and  the
vicinity of transition, while there are expectable deviations in
the localized phase: they are related with the fact that
theoretical results correspond to  $\ln\langle g \rangle$, while
numerical to  $\langle \ln g \rangle$. According to  (125),
the difference of two situations corresponds to
redefinition of  $\xi$ by the constant factor, which reduces
to the parallel shift in the logarithmic scale of Fig.\,15,a.

Comparison of the same dependence with the physical experiment
\cite{49} is possible under assumption that  $L$ is
replaced by the length  $L_{in}\propto T^{-\alpha}$,
characterizing the inelastic processes. Unfortunately, there are
no grounds for $\alpha$ to be the same in the metallic
and localized phase, and it can have a slow drift as a function
of disorder. In Fig.\,15,b it was suggested
that $\alpha$ is the piecewise constant quantity,
taking different values in the metal and insulator.  Such
assumption does not strongly affect the results in the critical
region where the length dependence of  $g_L$ is rather slow.  The
latter region is poorly presented in Fig.\,15,b, and in fact it
illustrates a situation not very close to the critical point. On
the other hand, the critical behavior obtained in \cite{49}, is
excellently described by the Vollhardt and W$\ddot o$lfle theory:
the values  $s=1.0\pm 0.1$ for the conductivity  exponent
and $z=2.94\pm 0.3$ for the dynamical exponent
agree with the theoretical  results $s=1$ and $z=3$.
\begin{figure*}
\centerline{\includegraphics[width=5.1 in]{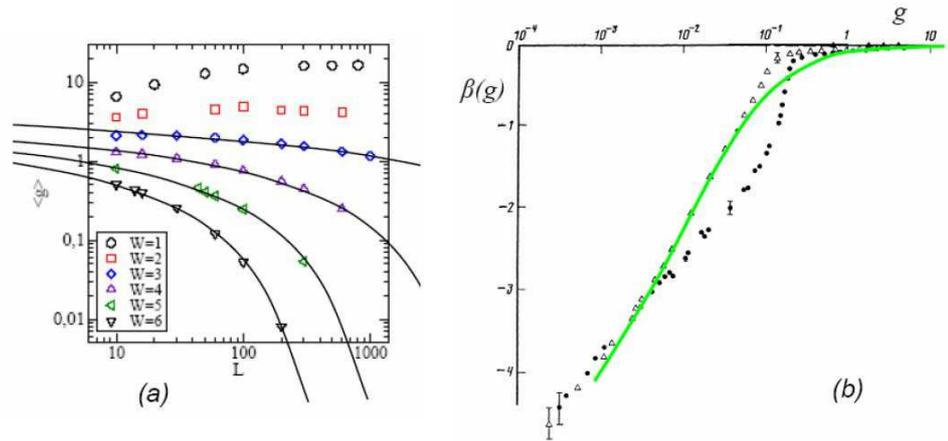}}
\caption{\footnotesize (a) Comparison of theoretical scaling
curves for $d=2$ with numerical results by Markos
\cite[Fig.37]{29}. The form of the curves was determined without
adjustable parameters, only parallel shifts along two axes were
made. (b) Comparison of the theoretical $\beta$ function for $d=2$
(Fig.5) with the empirical  $\beta$ function extracted from
experiment \cite{211} under assumption $L\propto T^{-\alpha}$. The
open and dark symbols corresponds to the freshly cleaved  surface
of Ge and to bicrystals of Ge.} \label{fig16}
\end{figure*}

The length dependence of $g_L$  for  $d=2$ can be
compared (Fig.\,16,a) with the numerical data by Mar-
kos \cite{29}.  There is a good agreement in the region
$W>2$, while for weak disorder numerical results display a strong
violation of scaling: it can be related with the protracted
ballistic regime  \cite{29} or existence of the 2D
metal-insulator transition.  The latter, according to \cite{27},
occurs in roughly half cases and belongs to  the
Kosterlitz--Thouless type, compatible with
scaling theory  \cite{1}. From viewpoint of  the general
analysis \cite{33}, a situation at $d=2$ is special and
probably reduces to the Vollhardt and W$\ddot o$lfle theory
not in all  cases.

The theoretical $\beta$ function for $d=2$ (Fig.\,5) can be
compared with  the empirical $\beta$ function (Fig.\,16,b)
obtained by Zavaritskaya \cite{211} under assumption  $L\propto
T^{-\alpha}$. The use of the constant $\alpha$ allows to describe
a situation both for large and small $g$. Some deviations are
present only in the region $g\sim 0.1$, where the experimental
results are also ambiguous.

\begin{center}
{\bf 8. SITUATION IN HIGHER DIMENSIONS} \end{center}

\begin{center}
{\bf 8.1. Dimensions $d>4$}
\end{center}

For  $d> 4$ we have for  $I_2(m)$ in (28)
$$
I_2(m) = -m^2 \left\{
c_{d}\Lambda^{d-4} +
O \left( m^{d-4}, L^{4-d} \right)
\right\}\,,
\eqno(126)
$$
i.e. analytical calculation is possible (in the main
approximation) for arbitrary values  of $m$ and $L^{-1}$.
Indeed, for  $m\agt L^{-1}$ the sum can be estimated by the
integral, which converges at the lower limit already for
$m=0$, so finiteness of $m$ gives only small corrections.
In the case  $m\alt L^{-1}$, the main effect of finite
$L$ is related with absence of the term with  ${ q}=0$,
which can be estimated as restriction  $ |{ q}|\agt L^{-1}$
in the integral approximation.

Substitution to
(13) reveals the possibility to neglect  $b_1$ and leads to the
scaling relation
$$
\pm \frac{1}{x^2} = y^2 -\frac{1}{y^2}
\eqno(127)
$$
in variables
$$
y=\frac{\xi_{0D}}{L} \left(\frac{a}{L} \right)^{(d-4)/4}\,,\qquad
x=\frac{\xi}{L} \left(\frac{a}{L} \right)^{(d-4)/4}\,,
\eqno(128)
$$
where we redefined the scales of  $\xi_{0D}$ and  $\xi$, in
order to obtain the unit coefficients in (127).
Scaling relations  (127,\,128) contain the atomic scale
$a$ due to nonrenormalizability of theory
\cite{28}. The critical point corresponds to  $y=1$, so
that
$$
\frac{\xi_{0D}}{L} \sim \left(\frac{L}{a} \right)^{(d-4)/4}
\,,\qquad \tau=0
\eqno(129)
$$
and the small $z$ asymptotics can be used for $H_T(z)$
(see Eq.65).
In  the vicinity of transition we can replace  $\xi_{0D}/L$ by
$\sqrt{g_L}$ and equations (127,\,128) determine the
length dependence  of  $g_L$; in particular, at the
transition point
$$
g_L \sim \left(\frac{L}{a} \right)^{(d-4)/2}
\,,\qquad \tau=0 \,.
\eqno(130)
$$
The physical sense of this result is clarified by the fact,
that it can be obtained from the self-consistency equation
for an infinite system (following from relations of Sec.2)
$$
D(\omega) = A\tau +B \left[\frac{-i\omega}{D(\omega)}
\right]^{1/2\nu}
\eqno(131)
$$
if one replace  $D(\omega)\to D_L$, $-i\omega\to\gamma$
and accept $\gamma$ (for $\tau=0$) to be of the order of level
spacing $\Delta\propto L^{-d}$. For $d<4$  it gives
$g_L=const$, while for  $d>4$  it reduces to  (130).

\begin{center}
{\bf 8.2. Four-dimensional case}
\end{center}

For $d=4$ we have analogously
$$
I_2(m) = \left\{
\begin{array}{cc}
{ \displaystyle  -c_4\, m^2
\ln\frac{\Lambda}{m} +
O \left( 1 \right)
\,,}\qquad & mL\agt 1 \\
{      }\\
{ \displaystyle  - c_4\, m^2
\ln(\Lambda L) +
O \left( 1 \right)
}\,,\qquad
 & mL\alt 1
\end{array}  \right. \,,
\eqno(132)
$$
and two results differ by $\ln(mL)$, which in the actual
region reduces to the double logarithmic quantity. Neglecting
 such quantities, we can obtain the scaling relation  (127) in
variables
$$
y=\frac{\xi_{0D}}{ L} \left[\ln(L/a)\right]^{-1/4}
\,,\qquad
x=\frac{\xi}{L}  \frac{\left[\ln(L/a)\right]^{1/4}}
{\left[\ln(\xi/a)\right]^{1/2}}             \,.
\eqno(133)
$$
In the vicinity of transition we can replace  $\xi_{0D}/L$
by $\sqrt{g_L}$ and obtain
$$
g_L \sim \left[\ln(L/a)\right]^{(d-4)/2}
\,,\qquad \tau=0 \,.
\eqno(134)
$$
As was explained in  \cite{28}, Eqs.127,\,133 allow to produce the
usual constructions of scaling curves, if the quantity $y$ is
considered as a function of the "modified length" $\mu(L)=L
\left[\ln(L/a)\right]^{-1/4}$; then a change of the scale for
$\mu(L)$ allows to reduce all dependencies for  $\tau>0$ and
$\tau<0$ to two universal curves. It should be emphasized, that
the critical point cannot be determined by the condition
$g_L=const$.

\begin{center}
{\bf 9. CONCLUSION}
\end{center}

The present paper continues the line initiated by the
previous publication \cite{28}:  since there are serious
indications  \cite{33,41} that the Vollhardt and W$\ddot o$lfle
theory predicts the correct critical behavior, it is desirable
to derive its consequences (as many as possible)  and
compare them with the numerical and physical experiments on the
level of raw data. Such approach already proved its value
\cite{28}: the results  $\nu=1.3\div 1.6$,
obtained usually for $d=3$ in numerical papers, can be
explained by  the fact that dependence $L+L_0$ with $L_0>0$
is interpreted as  $L^{1/\nu}$ with $\nu>1$, while the raw
data are excellently compatible with the self-consistent
theory. The finite-size scaling relations for the conductance and
Gell-Mann -- Low functions $\beta(g)$ obtained in the present
paper are also in a good agreement with numerical and physical
experiments.

In the present paper, we have elaborated a new definition for
the conductance of finite systems, brightening the questions
formulated in Sec.1.   It appears, that both self-consistent
theory of localization and the quantum-mechanical analysis based
on the shell model lead to the same definition, closely related
with definition by Thouless. It gives one more serious argument
in favour of the Vollhardt and W$\ddot o$lfle theory. Expansion
of the $\beta$ function in  $1/g$ shows that there are no
contradictions on the perturbative level with the results
of the $\sigma$ model approach in two dimensions.
 Further, in the case of validity of self-consistent
theory, the formalism of
dimensional regularization is incompatible with the physical
essence of the problem. Probably, it is the reason both for
the high-gradient catastrophe, and contradiction with
the Vollhardt and W$\ddot o$lfle theory in the space
dimension $d=2+\epsilon$.

The new definition will probably resolve the
problem of pathological singularities in the conductance
distribution \cite{29}, which cannot exist in finite systems.
Their observation in numerical studies  \cite{29} is  probably
explained by the fact the considered system was not sufficiently
isolated from environment and the thermodynamic limit $L\to\infty$
was effectively taken along one of  the
coordinate axes.

The above approach suggests the simple argumentation
on the spatial dispersion of the diffusion coefficient.
It is easily proved for the localized phase \cite{33}, that
$D(\omega, q)$ has a regular expansion in  $q^2$. However, it
does not exclude the appearance  of non-integer powers of
$q$ at the critical point  \cite{46} due to possibility of
constructions
$$
D(\omega, q) \sim \left( 1+\xi^2 q^2 \right)^\eta   \,,
\eqno(135)
$$
becoming singular in the $\xi\to\infty$ limit.
In a finite system the role of  $\xi$ is played by
$\xi_{0D}$, which is regular
at the transition point; with such
replacement, Eq.135 is valid in the metallic phase. However,
the absence of such dispersion in the metallic regime is
easily established from the kinetic equation. According to
 \cite{33}, the following result is valid instead (135)
$$
D(\omega, q) =(-i\omega) \xi^2  \left( 1+d_1 q^2 +
d_2 q^4 +\ldots \right)
\eqno(136)
$$
(with $d_i$ independent of $\tau$), which
reveals no pathologies for  replacement of  $\xi$ by
$\xi_{0D}$.
It should be noted that
 Wegner's exact result  $D(\omega,0)\sim \omega^{(d-2)/d}$
for the critical point \cite{47}, following from
(131) for $\tau=0$, cannot be
obtained in the case of essential spatial dispersion of
$D(\omega, q)$.  It is interesting, that the recent experiments
on the spreading of wave packet  \cite{48} are in  agreement with
the self-consistent theory and give no evidence of the anomalous
spatial dispersion.

The localization law for conductivity
$\sigma(\omega)\propto -i\omega$ was predicted almost 40
years ago \cite{49} but
has never been
observed experimentally.
The above analysis clarifies that its observation
is possible in
the closed systems under approximately the same
conditions as for existence of the persistent current in the
Aharonov--Bohm geometry (Fig.\,3) \cite{50,51,52}.

\begin{center}
{\it Appendix. Asymptotics of $g_L$ for $m L\to\infty$}
\end{center}

Consider the sum (27) for the Bloch boundary conditions in all
directions
$$
I(m) = \frac{1}{L^{d}} \sum\limits_{s_1,\ldots,s_d} \left.
\,\frac{1}{m^2 + q_1^2+\ldots+q_d^2}
\right|_{q_i=\frac{2\pi s_i+\varphi_i}{L}} \,,
\eqno(A.1)
$$
and introduce the so called $\alpha$-representation
$$
\frac{1}{m^2 + q^2} =\int_{0}^{\infty} \,d\alpha \,
e^{-\alpha\left(q^2+m^2\right)} \,.
\eqno(A.2)
$$
Then
$$
I(m) =\int_{0}^{\infty} \,d\alpha\,
e^{-\alpha m^2} \, \prod_{j=1}^{d} \, S_j(\alpha)\,,
$$
$$
S_j(\alpha)= L^{-1} \sum\limits_{s=-\infty}^{\infty} \left.
e^{-\alpha q_s^2} \right|_{q_s=\frac{2\pi s+\varphi_j}{L}} \,
\eqno(A.3)
$$
and the use of the Poisson summation formula \cite{36}
transforms $S_j(\alpha)$ to
$$
S_j(\alpha)=\frac{1}{\sqrt{4\pi\alpha}}
\sum\limits_{k_j=-\infty}^{\infty}
e^{ik_j\varphi_j -\frac{k_j^2 L^2}{4\alpha}} \,.
\eqno(A.4)
$$
Then (A.3) takes a form
$$
I(m) =\int_{0}^{\infty} \,\frac{d\alpha}{(4\pi\alpha)^{d/2}}
e^{-\alpha m^2}
\sum\limits_{\vec k}
e^{i{\vec k}\cdot{\vec \varphi} -
\frac{|{\vec k}|^2 L^2}{4\alpha}}  \,,
\eqno(A.5)
$$
where a vector  ${\vec k}=(k_1,\ldots,k_d)$ is introduced
and ${\vec k}\cdot{\vec \varphi}=\sum_j k_j \varphi_j$.
The term with ${\vec k}=0$  is calculated exactly
and corresponds to the continual approximation. The
main effect of discreteness is determined by the terms
with $|{\vec k}|=1$, which can be calculated for $mL\gg 1$
in the saddle-point approximation.
Remaining only  these terms, one has
$$
I(m) =\,\frac{m^{d-2}}{(4\pi)^{d/2}} \left[
\Gamma\left({ 1-\frac{d}{2}}\right) +
\right.
$$
$$
\left.
 \sqrt{\pi}
\left(\frac{mL}{2} \right)^{(1-d)/2}
e^{-mL} \sum_{j=1}^{d} 2\cos{\varphi_j} \right] \,.
\eqno(A.6)
$$
Taking the difference of two such expressions with
$\varphi_1=0$ and  $\varphi_1=\pi$,
$$
\left. I(m)\right|_{\varphi_1=0} -
\left. I(m)\right|_{\varphi_1=\pi} =
$$
$$
=\,\frac{4\sqrt{\pi}}{(4\pi)^{d/2}}\,m^{d-2}
\left(\frac{mL}{2} \right)^{(1-d)/2} e^{-mL}
 \,,
\eqno(A.7)
$$
we come to  Eq.59.


\end{document}